\documentclass[12pt]{article}
\usepackage{a4wide,amssymb,cite}
\pdfoutput=1

\usepackage{a4wide,amssymb,graphicx}
\usepackage{epsfig}
\usepackage[usenames,dvipsnames]{color}
\usepackage{slashed}
\usepackage{amssymb,cite,graphicx}
\usepackage{slashed}
\usepackage{amsmath,bm,bbm}
\usepackage{amsfonts}
\usepackage[titletoc,title]{appendix}
\usepackage[small]{caption}
\usepackage[margin=1in]{geometry}
\usepackage[multiple]{footmisc}
\usepackage{mathtools}
\usepackage{slashed}
\usepackage[nottoc]{tocbibind}
\usepackage{xcolor}

\usepackage{tikz}
\usetikzlibrary{shapes,arrows,positioning,automata,backgrounds,calc,er,patterns}
\usepackage[compat=1.1.0]{tikz-feynman}
\usetikzlibrary{trees}
\usetikzlibrary{decorations.pathmorphing}
\usetikzlibrary{decorations.markings}

\newcommand{\be}{\begin{equation}}
\newcommand{\ee}{\end{equation}}
\newcommand{\bea}{\begin{eqnarray}}
\newcommand{\eea}{\end{eqnarray}}

\def\circa#1{\,\raise.3ex\hbox{$#1$\kern-.75em\lower1ex\hbox{$\sim$}}\,}

\begin{document}

\begin{titlepage}
%
%


%

\begin{centering}
\vspace{1cm}
{\Large {\bf Graceful exit from inflation and reheating \vspace{.2cm} \\  with twin waterfalls}} \\

\vspace{1.5cm}

{\bf Hyun Min Lee$^\dagger$ and Adriana Menkara$^\sharp$ }
\\
\vspace{.5cm}

{\it  Department of Physics, Chung-Ang University, Seoul 06974, Korea.}

\vspace{.5cm}


\end{centering}
\vspace{2cm}

\begin{abstract}
\noindent
We study the hybrid inflation with a pseudo-Nambu-Goldstone boson inflaton and two waterfall scalar fields. The $Z_2$ symmetry for the waterfall fields keeps inflaton potential flat against quantum corrections coming from the waterfall couplings, and it is broken spontaneously in the vacuum without a domain wall problem within the Hubble horizon of our universe. We show that the $Z_2$ invariant Higgs portal couplings to the waterfall fields are responsible for the reheating process, leading to a sufficiently large reheating temperature after inflation. In the presence of an extra $Z'_2$ symmetry, one of the waterfall fields or another singlet scalar field becomes a dark matter candidate. In particular, we find that preheating is sufficient to account for the correct relic density of the waterfall dark matter.

\end{abstract}

\vspace{3cm}

\begin{flushleft} 
$^\dagger$Email: hminlee@cau.ac.kr \\
$^\sharp$Email: amenkara@cau.ac.kr 
\end{flushleft}

\end{titlepage}

\section{Introduction}

Cosmic inflation \cite{guth} has been successful for solving various problems in the Standard Big Bang cosmology, such as the initial conditions for the Friedman-Lemaitre-Robertson-Walker universe (namely, homogeneity, isotropy, flatness), the inhomogeneities imprinted in the Cosmic Microwave Background (CMB) anisotropies \cite{planck,keck}, the large scale structures, etc. The period of an exponential expansion of the universe requires a very flat potential of the scalar field, the so called inflaton, thus accounting for the early vacuum energy domination. 

On the other hand, the exponential expansion must end in order to recover the success of the Big Bang cosmology, namely, the Big Bang Nucleosynthesis (BBN). Otherwise, the universe would have continued to undergo the exponential expansion, being left with nothing. This requires the graceful exit from inflation \cite{chaotic} and the reheating process \cite{reheating0}. Thus, we need to specify the interactions between the inflaton and the SM in order to populate the Standard Model particles in the post-inflation regime.

In chaotic inflation models with a single inflaton \cite{chaotic}, inflation ends due to the violation of the slow-roll condition, so the graceful exit from inflation is naturally realized. Hybrid inflation \cite{hybrid}, on the other hand, requires at least two scalar fields for inflation and graceful exit at the same time. In this case, the inflaton drives a slow-roll inflation whereas the graceful exit or the violation of the slow-roll condition is achieved due to the tachyonic instability of another scalar field \cite{PT}, the so called waterfall field. It is conceivable to have multiple scalar fields in an UV complete theory, such as in compactifications of string theory, so hybrid inflation scenarios are more realistic cases and even single-field inflation models can be regarded as a certain limit of decoupling heavy particles.

In this article, we consider a model for hybrid inflation and its reheating dynamics with a pseudo-Nambu-Goldstone boson(pNGB) inflaton and two waterfall scalar fields \cite{twin}.  In this model, the shift symmetry of the inflaton ensures a naturally flat potential for the inflaton, whereas the $Z_2$ discrete symmetry for two waterfall fields \cite{twin,sundrum} renders the inflaton potential insensitive to loop corrections coming from the couplings between the waterfall fields and the inflaton. 

From the post-inflationary dynamics of the hybrid inflation, we discuss the preheating effects in the presence of waterfall-field dependent masses and the perturbative decay of the waterfall field for reheating. The $Z_2$ invariant Higgs portal couplings to  the waterfall fields are responsible for reheating. Moreover, focusing on the case that  the twin waterfall field or another singlet scalar field coupled to the waterfall field is a stable dark matter candidate due to another $Z'_2$ symmetry, we also calculate the dark matter abundance from preheating and/or the perturbative decay of the oscillating waterfall field. There is one appendix dealing with the details of the particle production and preheating during the waterfall transition.

The paper is organized as follows. 
We first present the model setup for the hybrid inflation with a pNGB inflaton and two waterfall fields and the condition for the waterfall transition and the robustness of the tree-level inflaton potential in this case. 
Then, we show the inflationary predictions of the model, the waterfall field dynamics as well as the vacuum structure, constraining the parameters of the waterfall sector.  Next, we discuss the preheating effects from the waterfall transition and the reheating from the perturbative decay of the waterfall field after preheating. In the presence of an extra $Z'_2$ symmetry, we regard the twin waterfall field as a dark matter candidate and discuss the preheating and the perturbative decay of the waterfall field for dark matter production. Finally, conclusions are drawn.

\section{The model}

We consider a pNGB $\phi$ as the inflaton and two real scalar fields $\chi_1, \chi_2$ as the waterfall fields in the hybrid inflation scenarios \cite{twin}. We discuss the roles of discrete symmetries and their origin for the UV insensitive inflaton potential and the stability of dark matter.

\subsection{Hybrid inflation}

We decompose the scalar potential for the hybrid inflation into the following,
\bea
V(\phi,\chi_1,\chi_2,H) =V_I(\phi)+V_W(\phi,\chi_1,\chi_2) +V_{\rm RH}(\chi_1,\chi_2,H)
\eea
where $V_I(\phi)$ is the inflaton potential, $V_W(\phi,\chi_1,\chi_2)$ is the waterfall field part consisting of two waterfall fields, and
$V_{\rm RH}(\chi_1,\chi_2,H)$ is the potential part for reheating the universe by the couplings of the waterfall fields to the Standard Model Higgs $H$.

Imposing a $Z_2$ discrete symmetry \cite{sundrum,twin} with
\bea
Z_2:\quad \phi\rightarrow-\phi,  \qquad  \chi_1\leftrightarrow \chi_2,
\eea
we  take the scalar potential for the hybrid inflation in the following form,
\bea
V_I(\phi) &=& V_0 +A(\phi), \label{inflaton} \\
V_W(\phi,\chi_1,\chi_2)&=&B(\phi) (\chi^2_1-\chi^2_2) +\frac{1}{2} m^2_\chi (\chi^2_1+\chi^2_2) -
\alpha^2 \chi_1\chi_2\nonumber \\
&&+\beta (\chi^3_1+\chi^3_2) + \gamma (\chi^2_1\chi_2+\chi_1\chi^2_2) \nonumber \\
&&+\frac{1}{4} \lambda_\chi (\chi^4_1+\chi^4_2) + \frac{1}{2} {\bar\lambda}_\chi \chi^2_1 \chi^2_2+ \frac{1}{3} \lambda'_\chi(\chi^3_1\chi_2+\chi_1\chi^3_2),  \label{full}
\eea
and 
\bea
V_{\rm RH}(\chi_1,\chi_2,H)= \kappa_1 (\chi^2_1+\chi^2_2)|H|^2 +\kappa_2 \chi_1\chi_2 |H|^2.
\eea
Here, $V_0$ is the constant vacuum energy during inflation, and $A(\phi), B(\phi)$ are arbitrary functions of $\phi$, satisfying $A(-\phi)=A(\phi)$ and $B(-\phi)=-B(\phi)$. 
The simple choices for $A(\phi), B(\phi)$, are $A(\phi)=-\frac{1}{2}m^2_\phi \phi^2$ and $B(\phi)=-g\phi$.
The renormalizable waterfall field couplings, $\phi^2 (\chi^2_1+\chi^2_2)$, can be also introduced, being consistent with the $Z_2$ symmetry and contributing to the effective masses for the waterfall fields during inflation. But, as will be shown below, the shift symmetry for $\phi$ makes such higher order terms in $\phi$ naturally suppressed.  

In the presence of a separate $Z'_2$ symmetry acting only on $\chi_2$, 
\bea
Z'_2:\quad \phi\to\phi,  \qquad  \chi_1\to \chi_1, \qquad \chi_2\to -\chi_2, 
\eea
we can set $\alpha=\beta=\gamma=\lambda'_\chi=0$ in combination with the $Z_2$ symmetry, and $\kappa_2=0$, so $\chi_2$ can be a candidate for dark matter. In this case, the independent parameters of the waterfall fields are reduced to $\lambda_\chi, {\bar\lambda}_\chi $, and $\kappa_1$ only.

For the concrete discussion on inflation and waterfall transition, in the later analysis, we take the inflaton potential in eq.~(\ref{inflaton}) and the waterfall field couplings for the inflaton in eq.~(\ref{full}) in the following periodic forms \cite{twin},
\bea
A(\phi)&=&\Lambda^4 \cos\Big(\frac{\phi}{f}\Big), \label{fA} \\
B(\phi)&=&-\frac{1}{2} \mu^2 \sin\Big(\frac{\phi}{2f} \Big). \label{fB}
\eea
Then, the shift symmetry for $\phi$ is broken into a discrete one, $\phi\to \phi+4\pi f$.
Expanding the sinuous functions around the origin, we can have $A(\phi)\simeq \Lambda^4-\frac{1}{2} m^2_\phi \phi^2$ with $m^2_\phi=\frac{\Lambda^4}{f^2}$ and $B(\phi)\simeq -g\phi$ with $g=\frac{\mu^2}{4f}$.  For hybrid inflation, we need to choose  $V_0\gtrsim \Lambda^4$, so the graceful exit from inflation is possible due to the transition with waterfall fields.

We first identify the inflaton-dependent mass eigenvalues for the waterfall fields  by
\bea
m^2_1(\phi) &=& m^2_\chi - \sqrt{ 4B^2(\phi)+\alpha^4},  \label{chi1mass0}\\
m^2_2(\phi)&=& m^2_\chi +\sqrt{4B^2(\phi)+\alpha^4}, \label{chi2mass0}
\eea
and the mixing angle $\theta$ between the waterfall fields depends on the inflaton field by
\bea
\sin2\theta(\phi)=\frac{2\alpha^2}{m^2_2(\phi)-m^2_1(\phi)}.
\eea
Here, we can keep the kinetic terms for the waterfall fields in the approximately canonical forms during the slow-roll inflation.
For $\alpha=0$, there is no mixing between the waterfall fields, so we can just keep track of the waterfall field $\chi_1$ to determine the end of inflation. 

During inflation, there is no VEV for the waterfall fields for $m^2_1(\phi)>0$ and $m^2_2(\phi)>0$.  Namely, for $\phi<\phi_c$ where $\phi_c$ satisfies $B^2(\phi_c)=m^4_\chi-\alpha^4$ with $\alpha<m_\chi$ and $B(\phi)$ is a monotonically increasing function of $\phi$ near $\phi_c$, so the slow-roll inflation takes place. For instance, from eq.~(\ref{chi1mass0}) with eq.~(\ref{fB}), setting $m^2_1(\phi_c)=0$, we find the point of the waterfall transition $\phi_c$ as
\bea
\phi_c=2f\arcsin\Big(\sqrt{m^4_\chi-\alpha^4}/\mu^2\Big) \label{end}
\eea
with $\sqrt{m^4_\chi-\alpha^4}<\mu^2$ and $\alpha<m_\chi$.
Then, the waterfall fields are heavy enough for $\mu, m_\chi \gg H_I$ with $H_I$ being the Hubble scale during inflation, so we can describe the slow-roll inflation by the inflaton potential given in eq.~(\ref{inflaton}) with eq.~(\ref{fA}).
At $\phi=\phi_c$ the waterfall field with mass $m_1$ starts becoming unstable, ending inflation even if the slow-roll condition for the inflaton direction is not violated.

\subsection{UV insensitive inflaton potential}

Due to the couplings of the waterfall fields to the inflaton, we compute the one-loop Coleman-Weinberg potential for the inflaton in cutoff regularization with cutoff scale $M_*$, as follows,
\bea
V_{\rm CW} &=&\frac{1}{64\pi^2}\sum_{i=1,2} \left[2m^2_{\chi_i}M^2_* -m^4_{\chi_i} \ln\bigg( \frac{e^{\frac{1}{2}}M^2_* }{m^2_{\chi_i}}\bigg)  \right] \nonumber \\
&\simeq&\frac{1}{16\pi^2}\,m^2_\chi M^2_* -\frac{1}{64\pi^2}\Big(m^4_\chi +4B^2(\phi)+\alpha^4\Big)\ln  \frac{M^2_*}{m^2_\chi}.  \label{CW}
\eea 
Then, the constant vacuum energy proportional to $M^2_*$ must be renormalized to get the desirable inflation energy. On the other hand, the quadratically divergent part of the inflaton potential is cancelled between the waterfall fields due to the $Z_2$ discrete symmetry, and the logarithmically divergent terms of the inflaton potential can be ignored during inflation as far as $B^2(\phi)\lesssim 16\pi^2 |A(\phi)|$ is satisfied.  For instance, for the periodic forms given in eqs.~(\ref{fA}) and (\ref{fB}), we can set the bound on the loop corrections by $\mu^2\lesssim 8\pi\Lambda^2$.

\subsection{Origin of discrete symmetries}

We comment upon the origin of the discrete symmetries for the inflaton and the waterfall fields. Suppose that a $U(1)$ global symmetry is broken to a $Z_4$ symmetry, under which the inflaton $\phi$ and a complex scalar field $\Phi$, transform by
\bea
Z_4:\quad \phi\to \phi, \qquad \Phi\to -i \Phi.
\eea
Moreover, we take the CP symmetry in the dark sector as
\bea
{\rm CP}:\quad \phi\to -\phi, \qquad \Phi\to \Phi^*.
\eea
As a result of combining $Z_4\times {\rm CP}$, we get
\bea
Z_4\times {\rm CP}:\quad \phi\to -\phi, \qquad \Phi\to i\Phi^*.
\eea
In this case, writing $\Phi=\frac{1}{\sqrt{2}}(\chi_1+i\chi_2)$, we can realize $\phi\to-\phi$ and $\chi_1\leftrightarrow \chi_2$ under $Z_4\times {\rm CP}$, as required for the $Z_2$ symmetry, thus providing the potential for hybrid inflation in our model.
On the other hand, the separate $Z'_2$ symmetry for $\chi_2$ corresponds to
\bea
Z'_2:\quad \phi\to \phi,\qquad \Phi\to \Phi^*.
\eea

\section{Inflation and waterfall field dynamics}

We now discuss the inflationary predictions of the pNGB inflation with twin waterfall fields.
Ignoring the classical dynamics of the waterfall fields during inflation, we focus on the slow-roll inflation and the condition for the waterfall transition.  Then, we show how the vacuum structure is connected to the inflation regime, constraining the waterfall sector parameters.

\subsection{Inflationary predictions}

From the inflaton potential given in eq.~(\ref{fA}), we first obtaine the slow-roll parameters for inflation as
\bea
\epsilon &=& \frac{M^2_P \Lambda^8 \sin^2(\phi/f)}{2f^2 (V_0+\Lambda^4\cos(\phi/f))^2}, \\
\eta &=& -\frac{M^2_P \Lambda^4 \cos(\phi/f)}{f^2 (V_0+\Lambda^4\cos(\phi/f))}.
\eea
The number of efoldings is also obtained as
\bea
N&=&\frac{1}{M_P} \int^{\phi_c}_{\phi_*} \frac{1}{\sqrt{2\epsilon}}\, d\phi  \nonumber \\
&=& \frac{f^2}{2M^2_P \Lambda^4}\, \bigg[V_0 \ln \Big(\tan^2\Big(\frac{\phi_c}{2f}\Big)\Big)+\Lambda^4\ln \Big(\sin^2\Big(\frac{\phi_c}{f}\Big)\Big) \bigg] -(\phi_c\to \phi_*) 
\eea
where $\phi_*,\phi_c$ are the inflaton field values at the horizon exit and at the end of inflation, respectively. 

For $V_0\gg \Lambda^4$ for hybrid inflation, the slow-roll parameters and the number of efoldings are approximated \cite{twin} to
\bea
\eta_* &\simeq& -\frac{M^2_P \Lambda^4}{f^2V_0 }\, \cos(\phi_*/f), \\
\epsilon_* &\simeq & \frac{M^2_P \Lambda^8}{2f^2V^2_0}\,  \sin^2(\phi_*/f), \\
N &\simeq&  \frac{f^2 V_0}{M^2_P \Lambda^4}\,\ln \Big(\frac{\tan(\phi_c/(2f))}{\tan(\phi_*/(2f))} \Big). \label{Nefold}
\eea
As a result, the spectral index and the tensor-to-scalar ratio can be determined by
\bea
n_s &=&1+2\eta_*-6\epsilon_*,  \label{ns}\\
r&=&16\epsilon_*. \label{r}
\eea
The CMB normalization, $A_s=\frac{1}{24\pi^2} \frac{V_0+\Lambda^4}{\epsilon_* M^4_P}\simeq 2.1\times 10^{-9}$, leads to
\bea
r=3.2\times 10^7\,\cdot\frac{V_0}{M^4_P}.  \label{cmb}
\eea
In order to get a spectral index which is consistent with Planck data, $n_s=0.967\pm 0.0037 $ \cite{planck}, we need to choose $2\eta_*\simeq -0.0033$ because  $ \epsilon_*\ll |\eta_*|$ in our case. The critical value $\phi_c$ of the inflaton should satisfy $ \phi_*\lesssim \phi_c\lesssim f$ for the number of efoldings, $N=50-60$, to solve the horizon problem. 
Moreover, the Planck bound on the tensor-to-scalar ratio, $r<0.036$ \cite{keck}, gives rise to the upper bound on $H_I<4.6 \times 10^{13}\,{\rm GeV}$.

In order to check the parameter space for inflation in our model, from $H^2_I\simeq V_0/(3M^2_P)$ and eqs.~(\ref{ns}), (\ref{Nefold}) and (\ref{cmb}), it is more convenient to choose the following parametrization,
\bea
n_s&\simeq& 1+2\eta_*\simeq 1- \frac{|m^2_\phi|}{3H^2}\, \cos(\phi_*/f), \quad |m^2_\phi|= \frac{\Lambda^4}{f^2}, \\
N&=& \frac{\cos(\phi_*/f)}{|\eta_*|}\,  \ln \Big(\frac{\tan(\phi_c/(2f))}{\tan(\phi_*/(2f))} \Big), \\
\frac{H_I}{f}&=&2.9\times 10^{-4}\,\Big|\eta_*  \tan(\phi_*/f)\Big|, \label{cmb2}
\eea
together with the condition determining the end of inflation in eq.~(\ref{end}).
From eq.~(\ref{cmb2}), we find that the axion decay constant and the Hubble scalar during inflation are correlated.
Taking $|\eta_*|=0.033/2$ to get the consistent spectral index and $\cos(\phi_*/f)=0.95$, we get $f\simeq 6.4\times 10^5 H_I$, which is much larger than the Hubble scale, so the global symmetry responsible for the PNG inflaton is broken during inflation.  
As we vary the Hubble scale during inflation \cite{twin}, we can adjust $f, |m^2_\phi|$ and $\mu\sim m_\chi$ to maintain the successful predictions for inflation while the waterfall fields remain safely decoupled during inflation.

\subsection{Waterfall field dynamics}

At the end of inflation, the waterfall field in the direction with $m^2_1<0$ starts rolling fast at $\phi=\phi_c$, developing a nonzero background field and providing an extra contribution to the effective inflaton mass.
The effective inflaton mass squared from $\frac{\partial^2 V}{\partial\phi^2}$ is given by
\bea
m^2_{\phi, {\rm eff}} =  -\frac{\Lambda^4}{f^2} \cos\Big(\frac{\langle\phi\rangle}{f}\Big) +\frac{\mu^2}{8f^2}( \langle\chi^2_1\rangle - \langle\chi^2_2\rangle)\sin\Big(\frac{\langle\phi\rangle}{2f}\Big) \label{effinfmass}
\eea
where $\langle\,\,\rangle$ denotes the background field values. Then,  the inflaton moves toward a stable minimum near $\phi/f= \pi$, which is the common minimum for the inflaton potential and the waterfall-induced potential.
On the other hand, the waterfall field masses are given by eqs.~(\ref{chi1mass}) and (\ref{chi2mass}) with $\phi>\phi_c$, which are of order $\mu\sim m_\chi$ even after inflation ends.

After inflation ends, we can describe the post-inflation dynamics by the following set of the Boltzmann equations for scalar fields and the radiation energy density $\rho_R$,
\bea
{\ddot \phi}+3H {\dot \phi}  &=&-\Gamma_\phi {\dot\phi} +\frac{\Lambda^4}{f} \sin\Big(\frac{\phi}{f}\Big)+ \frac{\mu^2}{4f} \cos\Big(\frac{\phi}{2f} \Big) (\chi^2_1-\chi^2_2),  \\
{\ddot\chi}_1 + 3H {\dot\chi}_1  &=&-\Gamma_{\chi_1} {\dot\chi}_1+\mu^2\sin\Big(\frac{\phi}{2f} \Big) \chi_1-m^2_\chi \chi_1-\alpha^2 \chi_2-\lambda_\chi \chi^3_1 -{\bar\lambda}_\chi \chi_1 \chi^2_2, \\
{\ddot\chi}_2 + 3H {\dot\chi}_2 &=&-\Gamma_{\chi_2} {\dot\chi}_2 -\mu^2\sin\Big(\frac{\phi}{2f} \Big) \chi_2-m^2_\chi \chi_2-\alpha^2\chi_1-\lambda_\chi \chi^3_2 -{\bar\lambda}_\chi \chi^2_1 \chi_2, \\
{\dot\rho}_R + 4H \rho_R &=&\Gamma_{\phi} {\dot\phi}^2 +\Gamma_{\chi_1}{\dot\chi}^2_1  +\Gamma_{\chi_2}{\dot\chi}^2_2 ,
\eea
and the Friedmann equation,
\bea
H^2= \frac{\rho_I+\rho_R}{3M^2_P},
\eea
where $\rho_I$ is the sum of energy densities for the inflaton and the waterfall fields, given by 
\bea
\rho_I = \frac{1}{2} {\dot\phi}^2 + \frac{1}{2} {\dot\chi}^2_1 + \frac{1}{2} {\dot\chi}^2_2 +V.
\eea
Here, we maintain the homogeneity for scalar fields in space, namely, we take the zero modes of scalar fields in momentum, but quantum fluctuations can give rise to nontrivial momentum modes of scalar fields during preheating or non-perturbative reheating, as will be discussed in the next section.

During the waterfall transition, the effective masses for the Higgs fields and the waterfall fields vary with time due to their couplings. We set the waterfall field $\chi_2$ to zero at the onset of the waterfall transition. Then, we find that the effective masses for the Higgs field and the waterfall field $\chi_2$, in the following simple form,
\bea
m^2_{H,{\rm eff}} &=&m^2_{H,0} +\kappa_1 \chi^2_1(t), \\
m^2_{\chi_2, {\rm eff}} &=& m^2_{\chi_2,0} +{\bar \lambda}_\chi \chi^2_1(t)
\eea
where $m^2_{H,0}, m^2_{\chi_2,0}$ are the squared bare masses, being independent of the field value of $\chi_1$.

\subsection{The vacuum structure}

We discuss the vacuum structure for the inflaton and the waterfall fields.

In the presence of a separate $Z'_2$ symmetry for $\chi_2$, which sets $\alpha=\beta=\gamma=\lambda'_\chi=0$ in the waterfall sector potential, the vacuum structure gets simplified due to the unique minimum of the potential at $\langle\phi\rangle=\pi f$,  $\langle\chi_1\rangle= v_\chi $ and  $\langle\chi_2\rangle=0$, with
\bea
 v_\chi= \sqrt{\frac{\mu^2-m^2_\chi}{\lambda_\chi}}\equiv \sqrt{\frac{m^2_1}{\lambda_\chi}}. \label{chivev}
\eea
Then, the cosmological constant during inflation can be fine-tuned  to the observed value in the true vacuum, as far as
\bea
V_0-\Lambda^4- \frac{1}{4}\lambda_\chi v^4_\chi \simeq 0, \label{zerocc}
\eea
thus constraining the parameters of the waterfall fields. We note than in the situation where there is no $Z'_2$ symmetry \cite{twin}, $\alpha\neq 0$ gives rise to a nonzero VEV for the waterfall field $\chi_2$, so, in general, the $Z_2$ symmetry is broken in the vacuum because $\langle\chi_1\rangle\neq \langle\chi_2\rangle$.  
Henceforth, we focus on the case with the $Z'_2$ symmetry for reheating and dark matter production. 

There might be a concern on the domain wall problem in our model, because the $Z_2$ symmetry for the waterfall fields is broken spontaneously in the vacuum. Namely, the vacuum $A$ with $\langle\phi\rangle=\pi f$,  $\langle\chi_1\rangle= v_\chi $ and  $\langle\chi_2\rangle=0$, is degenerate in energy with the vacuum $B$ with $\langle\phi\rangle=-\pi f$,  $\langle\chi_1\rangle= 0 $ and  $\langle\chi_2\rangle=v_\chi$. However, the inflaton direction is chosen to take positive values during the hybrid inflation and the universe evolves only to the vacuum $A$ through the waterfall transition as discussed above. Thus, domain walls could be formed in the entire space, but the regions evolving into the vacuum B are beyond the Hubble horizon during the hybrid inflation and remain so at present, so there is no observable signature of the domain walls in the current universe.  

We comment on the constraints on the parameters of the waterfall fields from the vacuum structure. 
For $V_0\gg \Lambda^4$, eqs.~(\ref{chivev}) and (\ref{zerocc}) give rise to $4V_0/\lambda_\chi= v^4_\chi= m^4_1/\lambda_\chi^2$. 
Then, the quartic coupling and the VEV for the waterfall fields are related to the dimensionful parameters of the inflation, as follows,
\bea
\lambda_\chi&=& \frac{m^4_1}{4V_0} = 1.4\times 10^{-20}\bigg(\frac{m_1}{100H_I}\bigg)^4 \bigg(\frac{H_I}{10^{5}\,{\rm GeV}}\bigg)^2,  \\
v_\chi&=& \sqrt{\frac{4V_0}{m^2_1}} = 0.035 M_P\bigg(\frac{100H_I}{m_1}\bigg). \label{VEV}
\eea 
Therefore, since $H_I\lesssim 1.6\times 10^{10}\,{\rm GeV}$ for $f\lesssim 10^{16}\,{\rm GeV}$, the waterfall self-coupling $\lambda_\chi$ is smaller than about $10^{-10}$ for $\mu\gtrsim 100 H_I$, and the VEV of the waterfall field $v_\chi$ is not far from the Planck scale.

Next, expanding around the VEV by $\phi=v_\phi+a$ and the waterfall fields as $\chi_1=v_\chi+{\tilde\chi}_1$, we obtain the inflaton mass and the mass eigenvalues for the waterfall fields in the true vacuum as
\bea
m^2_{a} &=& \frac{1}{f^2} \Big(\Lambda^4+\frac{1}{8} \mu^2 v^2_\chi\Big), \label{inflatonmass}  \\
m^2_{{\tilde\chi}_1}&=&  2\lambda_\chi v^2_\chi \nonumber \\
&=&2(\mu^2-m^2_\chi),  \label{chi1mass}  \\
m^2_{\chi_2} &=& \mu^2 + m^2_\chi +{\bar\lambda}_\chi v^2_\chi \nonumber  \\
&=& \mu^2+m^2_\chi +\frac{{\bar\lambda}_\chi }{\lambda_\chi}(\mu^2- m^2_\chi).  \label{chi2mass}
\eea
We also note that the vacuum stability for the waterfall field potential requires $\lambda_\chi>0$ and $\lambda_\chi+{\bar\lambda}_\chi>0$ for ${\bar\lambda}_\chi<0$. 

We find that the inflaton mass receives a tree-level correction due to the waterfall field coupling.
Using $|\eta_*|=\Lambda^4\cos(\phi_*/f)/(f^2 H^2_I)=0.033/2$ with $\cos(\phi_*/f)=0.95$ at horizon exit, $f=6.4\times 10^5\,H_I$, and $r\equiv \mu/m_\chi\gtrsim 1$, we can rewrite the effective inflaton mass as 
\bea
m^2_a=0.0495\, \frac{H^2_I}{\cos(\phi_*/f)} +(3\times 10^{-9})\Big( \frac{r^2}{r^2-1}\Big)\, \Big(\frac{m_1}{100H_I}\Big)^2\,v^2_\chi.
\eea
Thus, as compared to the tachyonic mass for the waterfall field, we recall $m^2_1=\lambda_\chi v^2_\chi$ with $\lambda_\chi\lesssim 10^{-10}$, so the effective inflaton mass is much heavier than that of the waterfall fields. Therefore, the inflaton settles down to the minimum of the potential rapidly, so it does not influence the reheating dynamics with the waterfall fields. It is also possible to reheat the universe from the inflaton energy density in the presence of the interactions between the inflaton and the SM particles, but reheating is dominated by the waterfall fields because $V_0\gg \Lambda^4$.

For $m_{{\tilde\chi}_1}<2m_{\chi_2}$, namely, for $(1+2\alpha_X)(\mu/m_\chi)^2+3-2\alpha_X<0$ with $\alpha_X={\bar\lambda}_\chi/\lambda_\chi$, the waterfall field  ${\tilde\chi}_1$ could not decay into a $\chi_2$ pair. This is the case in most of the parameter space where $\mu>m_\chi$ (waterfall condition) and $\alpha_X>-1$ (vacuum stability), so dark matter $\chi_2$ is not produced from the decay of the waterfall field $\chi_1$. However, as will be shown in the later section, dark matter $\chi_2$ can be still produced abundantly during preheating. 

Moreover, we find  the leading interaction terms between the inflaton $a$, the mass eigenstates of the waterfall fields, $ {\hat\chi}_{1,2}$, and the Higgs boson $h$, as follows,
\bea
{\cal L}_{\rm int} = \frac{\mu^2}{8f^2}\,v_\chi a^2 {\tilde\chi}_1+ \frac{\mu^2}{16f^2} a^2 ({\tilde\chi}^2_1-{\chi}^2_2)-{\bar\lambda}_\chi v_\chi {\tilde\chi}_1 \chi^2_2 -\kappa_1 v_\chi {\tilde\chi}_1 h^2+\cdots.\label{inflaton-int}
\eea
Thus, there is no quadratic divergence in the radiative corrections to the inflaton mass, due to the  $Z_2$ symmetry, although the radiative corrections to the inflaton mass are logarithmically divergent due to the cubic interactions in the first line in eq.~(\ref{inflaton-int}). We note that the decay modes, ${\tilde\chi}_1\to aa, \chi_2\chi_2$, are kinematically blocked, but ${\tilde\chi}_1\to hh$ is open for reheating.

\section{Reheating and dark matter production}

We first discuss the preheating effects from the waterfall transition and the reheating from the perturbative decay of the waterfall field. Then, we show the time evolution of radiation energy density and determine  the maximum temperature and the reheating temperature. We apply the results for obtaining the relic density for dark matter.

\subsection{Preheating}

After the end of inflation, the equations of motion for scalars, $\psi=H, \chi_2$, can be written in terms of the rescaled field, $\varphi=a\,\psi$, as follows,
\bea
\varphi^{\prime\prime} -\partial^2_i\varphi+\Big(m^2_{\psi,{\rm eff}} a^2+(6\zeta_\psi-1)\,\frac{a^{\prime\prime}}{a}  \Big)\varphi=0 \label{spert}
\eea
where $\zeta_\psi$ is the non-minimal coupling of the scalar field  $\psi$ and we define the conformal time $\eta=\int dt/a(t)$, with $a(t)$ being the scale factor. 
In the case with $\zeta_\psi=\frac{1}{6}$, we get the total number density of produced particles at the end of the waterfall transition by
\bea
n_\psi= \frac{1}{2\pi^2}\, \int^\infty_0 dk\, k^2 n^\psi_k, \label{numberd}
\eea
where
\bea
n^\psi_k =- \frac{g_\psi}{2} +\frac{1}{2\omega_k} \Big(|v'_k|^2 +\omega^2_k |v_k|^2 \Big), \quad \omega^2_k=k^2+m^2_{\psi,{\rm eff}}.
\label{modenumber}
\eea
Here, $v_k$ is the mode function for the scalar field, 
and $g_{\psi}$ is the number of degrees of freedom for $\psi=H,\chi_2$.

During the waterfall transition, we take the effective scalar masses \cite{Garcia-Bellido:2001dqy} as
\bea
m^2_{\psi,{\rm eff}}(t) = m^2_{\psi,0}+ \frac{1}{4} c_\psi v^2_\chi (1+\tanh \lambda(t-t_*))^2 \label{effmass}
\eea
where $m^2_{\psi,0}$ are the bare masses and $c_\psi$ are the couplings of the waterfall field $\chi_1$, given by $c_\psi=\kappa_1, {\bar\lambda}_\chi$ for $\psi=H, \chi_2$, respectively. Here, $\lambda$ is the strength of the waterfall transition, given by $\lambda=m_1/2$, with $m_1=\sqrt{\lambda_\chi} v_\chi$ being the tachyonic mass of the waterfall field $\chi_1$. $t_*\simeq \ln(32\pi^2/\lambda_\chi)/(2m_1)$ is the duration time of the waterfall transition \cite{Garcia-Bellido:2001dqy}, which is given by $t_*\simeq 14/m_1-26/m_1$ for $\lambda_\chi=10^{-20}-10^{-10}$. As $m_1\gg H_{\rm end}$, the waterfall transition takes place less than one Hubble time and finishes rapidly. Thus, the Hubble expansion can be ignored during the waterfall transition, allowing use to set the scale factor to $a=1$, 

From the results in eq.~(\ref{beta}) in the Appendix, we obtain the number density of produced particles with momentum $k$ by
\bea
n^\psi_k &=&\frac{ \cosh\Big(\pi (w_{\psi,2}-w_{\psi,1})/\lambda\Big)+\cosh(2\pi\delta) }{ 2\sinh(\pi \omega_{\psi,1}/\lambda)\sinh(\pi \omega_{\psi,2}/\lambda)} \label{npsi}
\eea
where $w^2_{\psi,1}=k^2+m^2_{\psi,0}$,  $w^2_{\psi,2}=k^2+m^2_{\psi,0}+c_\psi v^2_\chi$ and $\delta=\frac{1}{2}\sqrt{\frac{c_\psi v^2_\chi}{\lambda^2}-1}$.
Therefore, we obtain the energy density of the produced particles at the end of the waterfall transition by
\bea
\rho_\psi(t_*)=\frac{g_\psi}{2\pi^2}\, \int^\infty_0 dk\, k^2 n^\psi_k w_{\psi,2}.
\eea

As a consequence, the ratio of the produced energy density to the initial vacuum energy density, $V_0\simeq \frac{1}{4}\lambda_\chi v^4_\chi=\frac{m^4_1}{4\lambda_\chi}$, is given by
\bea
\frac{\rho_\psi(t_*)}{V_0} =\frac{2g_\psi\lambda_\chi}{\pi^2} \int^\infty_0 d\kappa \,\kappa^2 n^\psi_k  \sqrt{\kappa^2+\frac{m^2_{\psi,0}}{m^2_1}+\alpha_\psi} \label{analytic}
\eea
with $\kappa=k/m_1$ and $\alpha_\psi=c_\psi/\lambda_\chi$. 
In order to compare our results with the lattice calculations \cite{Garcia-Bellido:2001dqy}, for $m_{\psi,0}=0$, we quote the numerical solution for $\frac{\rho_\psi(t_*)}{V_0}$ in terms of a fitting function, $f(\alpha,\gamma)=\sqrt{\alpha+\gamma^2}-\gamma$ \cite{Garcia-Bellido:2001dqy}, as follows,
\bea
\frac{\rho_\psi(t_*)}{V_0} =2\times 10^{-3} g_\psi\lambda_\chi f(\alpha_\psi,1.3). \label{preheating}
\eea

Similarly, the numerical result for the number density at the waterfall transition is given  \cite{Garcia-Bellido:2001dqy} by 
\bea
n_\psi(t_*)=  10^{-3} g_\psi m^3_1f(\alpha_\psi,1.3)/\alpha_\psi.
\eea

\begin{figure}[!t]
\begin{center}
\includegraphics[width=0.45\textwidth,clip]{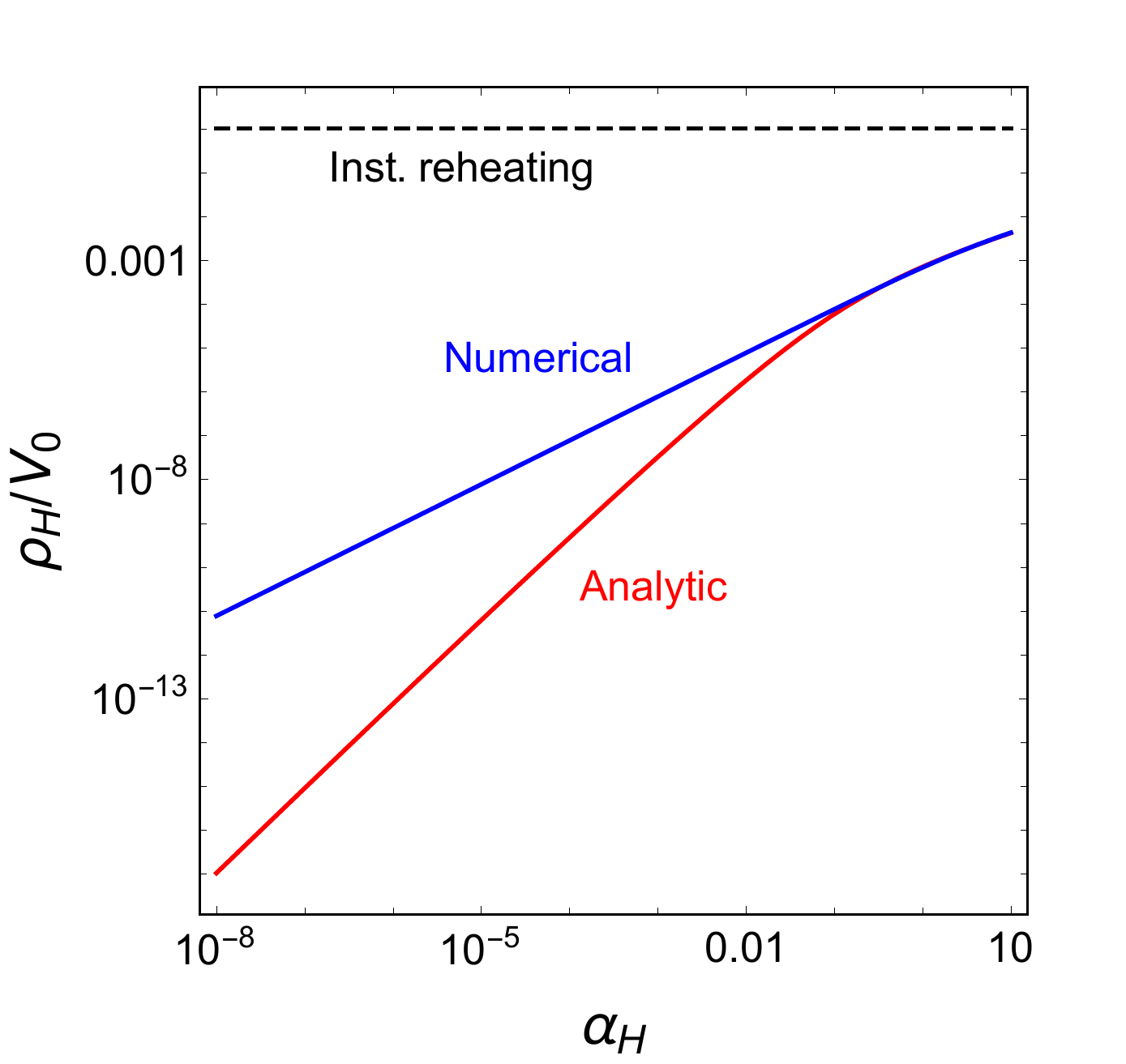}
\end{center}
\caption{Analytic and numerical solutions for $\rho_H/V_0$ from preheating as a function of $\alpha_H=\kappa_1/\lambda_\chi$ in red and blue lines, respectively. }
\label{fig:preheating}
\end{figure}

In Fig.~\ref{fig:preheating}, we depict the analytic and numerical results for $\rho_H/V_0 $ for the Higgs as a function of $\alpha_H=\kappa_1/\lambda_\chi$ in red and blue lines, respectively. We set the bare Higgs mass to $m_{H,0}=0$. 
The analytic result is based on the formula in eq.~(\ref{analytic}) while the numerical result is taken from eq.~(\ref{preheating}). For $\alpha_H\gtrsim 0.1$, both analytic and numerical results agree.  The black dashed line corresponds to the case for instantaneous reheating.

\subsection{Reheating completion}

At the end of preheating, the waterfall fields reach the minimum of the scalar potential and starts oscillating around it. 
Then, if preheating is not efficient, we can determine the reheating temperature from the perturbative decay of the waterfall field condensate $\chi_c$ by
\bea
T_{\rm RH}= \bigg(\frac{90}{\pi^2 g_{\rm RH}} \bigg)^{1/4} \sqrt{M_P \Gamma_{\chi_1}}  \label{RH}
\eea
where $g_{\rm RH}$ is the number of relativistic degrees of freedom at reheating completion and $ \Gamma_{\chi_1}$ is the decay rate for $\chi_1$.

Expanding the waterfall field as $\chi_1(t)= v_\chi+\chi_c(t)$ with $\chi_c(t)$ being the waterfall condensate, we get the total decay rate of the waterfall condensate  as
\bea
\Gamma_{\chi_1}= \Gamma_{\chi_1\to hh}+ \Gamma_{\chi_1\to\chi_2\chi_2}, \label{decay}
\eea
with
\bea
 \Gamma_{\chi_1\to hh}&=&\frac{\kappa_1^2 v^2_\chi }{2\pi m_{\chi_1}} \sqrt{1-\frac{4m^2_H}{m^2_{\chi_1}}}, \\
  \Gamma_{\chi_1\to\chi_2\chi_2} &=&\frac{{\bar\lambda}^2_\chi v^2_\chi }{2\pi m_{\chi_1}} \sqrt{1-\frac{4m^2_{\chi_2}}{m^2_{\chi_1}}}.
\eea
Then,  ignoring the Higgs and $\chi_2$ masses, we obtain the reheating temperature approximately by 
\bea
T_{\rm RH}\simeq \bigg(\frac{90}{\pi^2 g_{\rm RH}} \bigg)^{1/4} \bigg( \frac{\kappa^2_1+{\bar\lambda}^2_\chi}{4\pi \lambda_\chi}\bigg)^{1/2} \sqrt{M_P m_{\chi_1}}. 
\eea
Therefore, for $\Gamma_{\chi_1}\ll  H_I\sim m_{\chi_1}$, namely,  $\kappa^2_1+{\bar\lambda}^2_\chi\ll 4\pi \lambda_\chi$, and taking $H_I\lesssim 1.6\times 10^{10}\,{\rm GeV}$ for $f\lesssim 10^{16}\,{\rm GeV}$, we get the reheating temperature as $T_{\rm RH}\ll  10^{14}\, {\rm GeV}$.  

The waterfall transition takes place faster than the Hubble expansion, so preheating is instantaneous.
But, if the reheating during the waterfall oscillation is sizable and it is not instantaneous, the number of efoldings required to solve the horizon problem \cite{reheating1,HiggsR2} is modified to
\bea
N=61.1 +\Delta N -\ln \bigg(\frac{V^{1/4}_0}{H_k} \bigg) -\frac{1}{12} \ln \bigg(\frac{g_{\rm RH}}{106.75} \bigg)
\eea
where the correction to the number of efoldings due to the non-instantaneous reheating is given by
\bea
\Delta N= \frac{1}{12} \bigg(\frac{3w-1}{w+1} \bigg) \ln\bigg(\frac{45 \rho_{\chi_1}(t_*)}{\pi^2 g_{\rm RH} T^4_{\rm RH}} \bigg).\label{DN}
\eea
Here, $\rho_{\chi_1}(t_*)=V_0-\rho_R(t_*)$ is the energy density of the waterfall field at the end of the waterfall transition and $w$ is the equation of state for the waterfall condensate.
Here, $H_k$ is the Hubble parameter evaluated at the horizon exit for the Planck pivot scale, $k=0.05\,{\rm Mpc}^{-1}$, and $w$ is the averaged equation of state during reheating. 
Then, taking $w=0$ for $V_0\gtrsim \rho_R(t_*)$ and $g_{\rm RH}=106.75$ in eq.~(\ref{DN}), we obtain the number of efoldings  as
\bea
N=51.3-\frac{1}{3}\ln\bigg(\frac{H_I}{1.6\times 10^{10}\,{\rm GeV}}\bigg)-\frac{1}{12}\ln \bigg(1-\frac{\rho_R(t_*)}{V_0}\bigg)+\frac{1}{3}\ln\bigg(\frac{T_{\rm RH}}{10^{14}\,{\rm GeV}}\bigg).
\eea
As a result, we conclude that there is a wide range of the parameter space for a successful inflation, and a sufficiently large reheating temperature is achieved due to the decay of the waterfall field.

\subsection{Time evolution of radiation energy density}

We make a concrete discussion of the time evolution of the waterfall and radiation energy densities after preheating and determine the reheating temperature concretely. 

After preheating, we consider the Boltzmann equations in the presence of the coherent oscillation of the waterfall field, as follows,
\bea
{\dot\rho}_R + 4H \rho_R&=&\Gamma_{\chi_1\to hh}\rho_{\chi_1}, \\
{\dot\rho}_{\chi_1} + 3H \rho_{\chi_1} &=&-\Gamma_{\chi_1}\rho_{\chi_1},
\eea
with
\bea
H^2= \frac{\rho_R+\rho_{\chi_1}}{3M^2_P}. \label{Hubbleeq}
\eea
We take the initial condition for the waterfall energy density at preheating as $\rho_{\chi_1}(t_*)=V_0-\rho_R(t_*)$, where $\rho_R(t_*)=\rho_H$ is the radiation energy density from preheating. For a constant decay rate, $\Gamma_{\chi_1}\simeq \Gamma_{\chi_1\to hh}$, we obtain the analytic solutions to the above Boltzmann equations \cite{reheating} by
\bea
\rho_{\chi_1}(t) &=& \rho_{\chi_1}(t_*) \bigg(\frac{a(t)}{a_*}\bigg)^{-3}\, e^{-\Gamma_{\chi_1} (t-t_*)}, \\
\rho_R(t)&=& \rho_{\chi_1}(t_*) \bigg(\frac{a(t)}{a_*}\bigg)^{-4}\,\bigg[\frac{\rho_R(t_*)}{ \rho_{\chi_1}(t_*)}+\int^{u}_{u_*} \bigg(\frac{a(u)}{a_*}\bigg) e^{u_*-u} \,du \bigg]. \label{totalrad}
\eea 
Here, $u=\Gamma_{\chi_1}t$ and $u_*=\Gamma_{\chi_1}t_*$.

Combining eq.~(\ref{Hubbleeq}) and the effective continuity equation for the total energy density, $\rho=\rho_{\chi_1}+\rho_R$, given by
\bea
{\dot\rho} + 3H \rho (1+w(t))=0,
\eea
with $w(t)$ being the effective equation of state,  we find the total energy density as
\bea
\rho(t)=V_0 \bigg(1+\sqrt{\frac{3}{4} V_0}\, (1+{\overline w}) \,\frac{t-t_*}{M_P}\bigg)^{-2}.
\eea
where
\bea
{\overline w}(t) = \frac{1}{t-t_*} \int^t_{t_*} dt'\, w(t').
\eea
Here, we used $\rho_{\chi_1}+\rho_R\simeq V_0$ at the waterfall transition.  Then, for a slowly-varying ${\overline w}$, we also obtain the scale factor as a function of time,
\bea
\frac{a(t)}{a_*} \simeq \bigg( 1+\sqrt{\frac{3}{4} V_0} \, (1+{\overline w}) \,\frac{t-t_*}{M_P}\bigg)^{\frac{2}{3} \frac{1}{1+{\overline w}}}.
\label{scalefactor}
\eea

Setting ${\overline w}=0$ for the waterfall field oscillation near the quadratic potential and using the results in eqs.~(\ref{totalrad}) and (\ref{scalefactor}), we get the total radiation energy density after preheating, as follows,
\bea
\rho_R(t)= \rho_{\chi_1}(t_*) \bigg(1+\frac{v}{A}\bigg)^{-8/3}\,\bigg[ \frac{\rho_R(t_*)}{\rho_{\chi_1}(t_*)}+\int^v_0 \bigg(1+\frac{v'}{A}\bigg)^{2/3} e^{-v'} \,dv' \bigg] \label{rad}
\eea
where $v\equiv \Gamma_{\chi_1}(t-t_*)$ and
\bea
A&=&\frac{\Gamma_{\chi_1}}{m_{\chi_1}}\,\bigg(\frac{3}{4} \frac{V_0}{m^2_{\chi_1}M^2_P}\bigg)^{-1/2} \nonumber \\
&=& 4\sqrt{\frac{2}{3}} \frac{\Gamma_{\chi_1}}{ m_{\chi_1}} \bigg(\frac{v_\chi}{M_P} \bigg)^{-1}. \label{Apara}
\eea
Here, we used $V_0=\frac{m^4_1}{4\lambda_\chi}$ and $m^2_1=\lambda_\chi v^2_\chi=m^2_{\chi_1}/2$ in the second line of eq.~(\ref{Apara}).

\begin{figure}[!t]
\begin{center}
 \includegraphics[width=0.45\textwidth,clip]{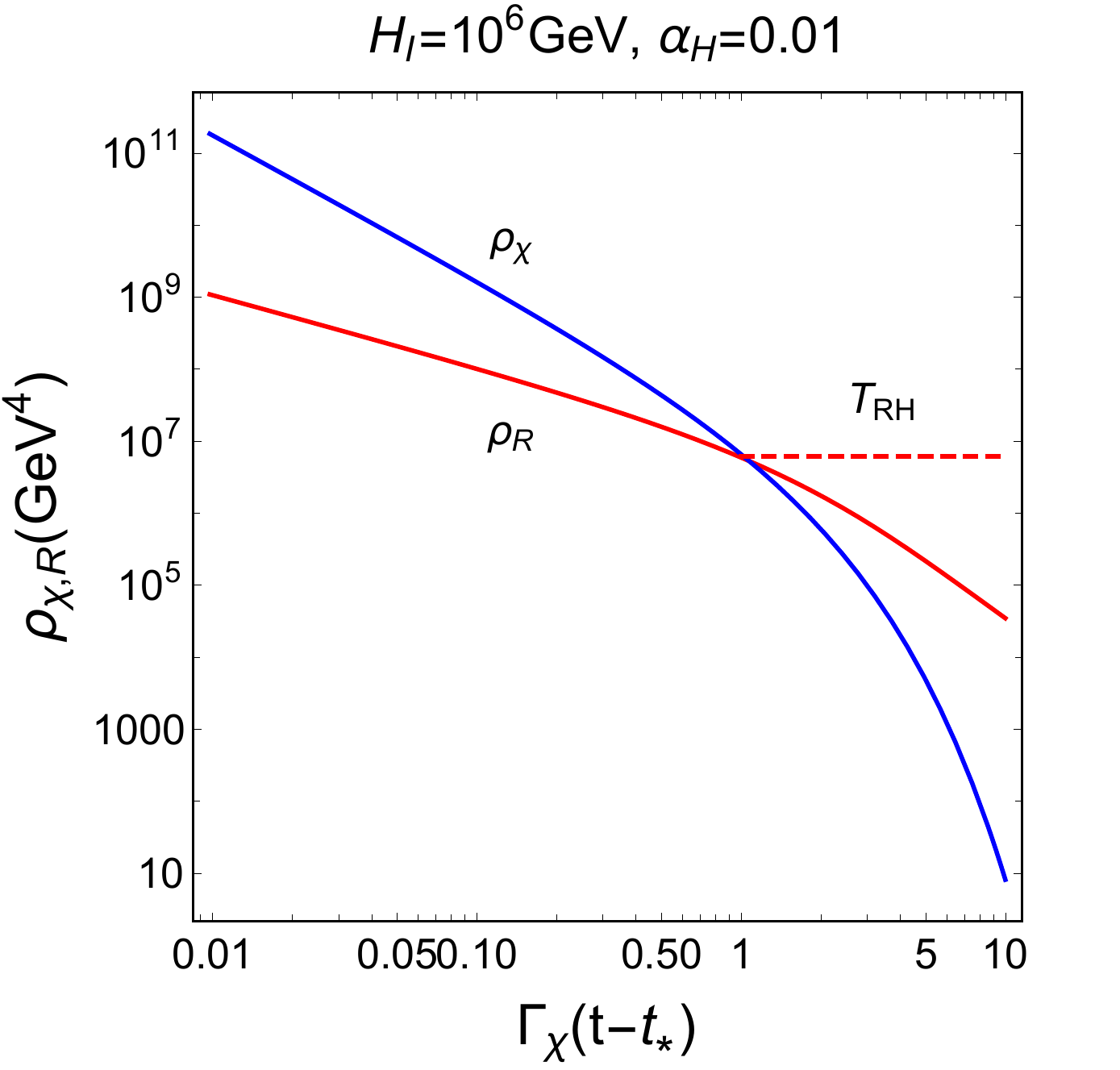}\,\,  \includegraphics[width=0.45\textwidth,clip]{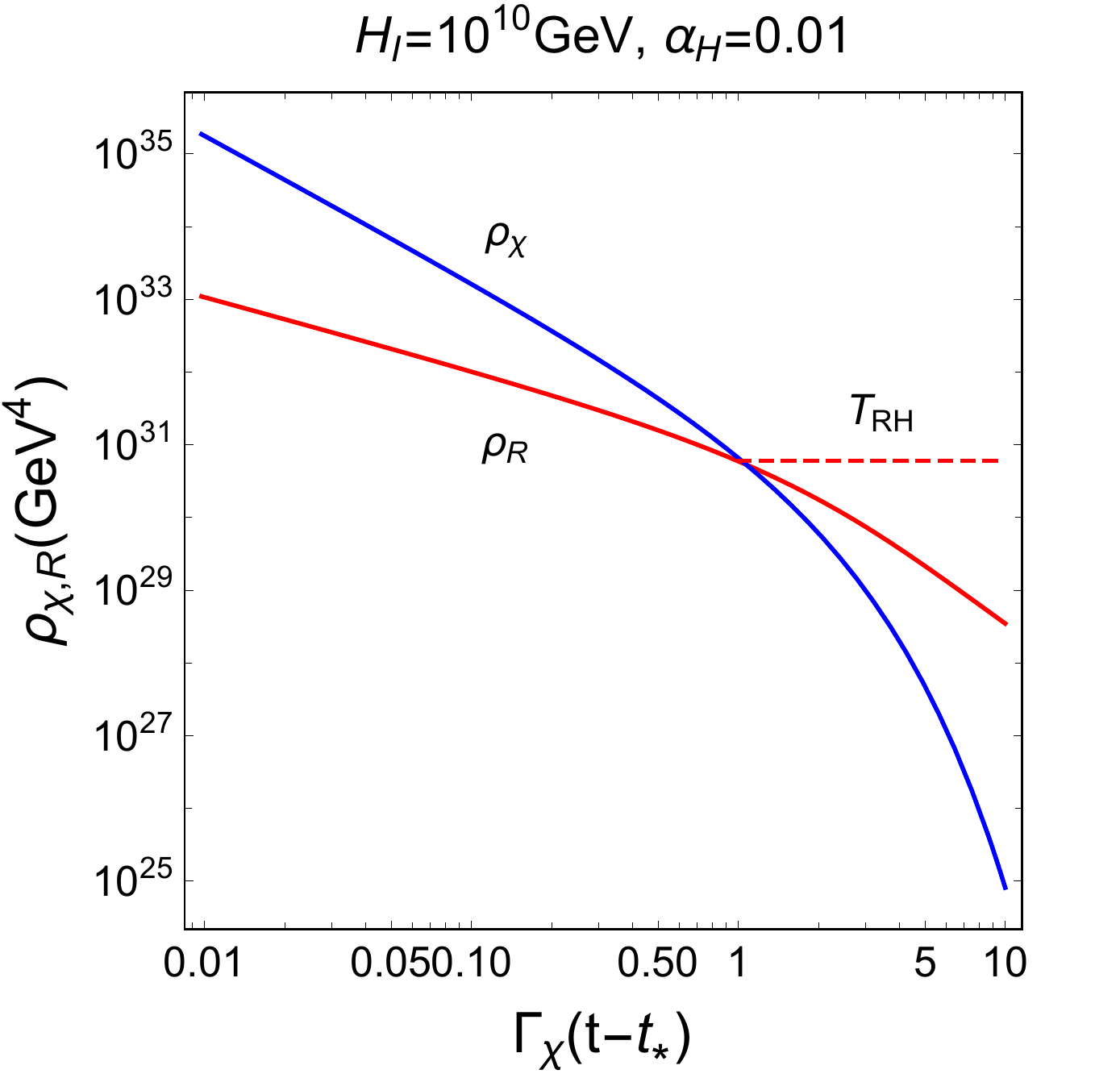}  
 \end{center}
\caption{$\rho_\chi, \rho_R$ as a function of $\Gamma_{\chi_1}(t-t_*)$ in blue and red solid lines, respectively. Red dashed lines correspond to the reheating completion. We took  $\alpha_H=0.01$ and $m_1=100\,{\rm H}_I$ for both plots, and $H_I=10^6,\, 10^{10}\,{\rm GeV}$ in left and right plots, respectively. }
\label{fig:evolution}
\end{figure}

In Fig.~\ref{fig:evolution}, we depict the time evolution of $\rho_{\chi_1}, \rho_R$ as a function of  $\Gamma_{\chi_1}(t-t_*)$ in blue and red solid lines, respectively, for $H_I=10^6\, 10^{10}\,{\rm GeV}$ in left and right plots. We find that the reheating temperature $T_{\rm RH}$ is determined from $\rho_R=\rho_{\chi_1}$, for which $\Gamma_\chi(t-t_*)\sim 1$. Thus, the results are consistent with the identification of the reheating temperature by eq.~(\ref{RH}).

The second term from reheating in eq.~(\ref{rad}) is maximized at $v_{\rm max}=\Gamma_{\chi_1}(t_{\rm max}-t_*)=0.80 A$, resulting in the maximum radiation energy,
\bea
\rho_R(t_{\rm max})= 1.8^{-8/3} \bigg(\rho_R(t_*)+1.0 A \rho_{\chi_1}(t_*) \bigg).
\eea  
Therefore, for $\rho_R(t_*)\lesssim V_0$ and $\rho_{\chi_1}(t_*)\simeq V_0$, we get the maximum ratio of reheating to  preheating contributions to the radiation energy density by
\bea
R_{\rm rh}\equiv \frac{1.0A\rho_{\chi_1}(t_*)}{\rho_R(t_*)}&\simeq& 1600\, \frac{\Gamma_{\chi_1}/m_{\chi_1}}{\lambda_\chi f(\alpha_H,1.3)}\,\bigg(\frac{v_\chi}{M_P}\bigg)^{-1} \nonumber \\
&\simeq& 130\,\frac{\alpha^2_H}{f(\alpha_H,1.3)}\,\bigg(\frac{v_\chi}{M_P}\bigg)^{-1}. \label{ratio}
\eea
Here, we used eq.~(\ref{Apara}) and the numerical result for preheating with the Higgs coupling, $\alpha_H=\kappa_1/\lambda_\chi$, in eq.~(\ref{preheating}), and took eq.~(\ref{decay}) in the second equality.
Thus, from eq.~(\ref{ratio}) with eq.~(\ref{VEV}), we find that 
\bea
R_{\rm rh}\simeq 3700\, \frac{\alpha^2_H}{f(\alpha_H,1.3)}\,\bigg(\frac{m_1}{100H_I}\bigg).
\eea

\begin{figure}[!t]
\begin{center}
 \includegraphics[width=0.45\textwidth,clip]{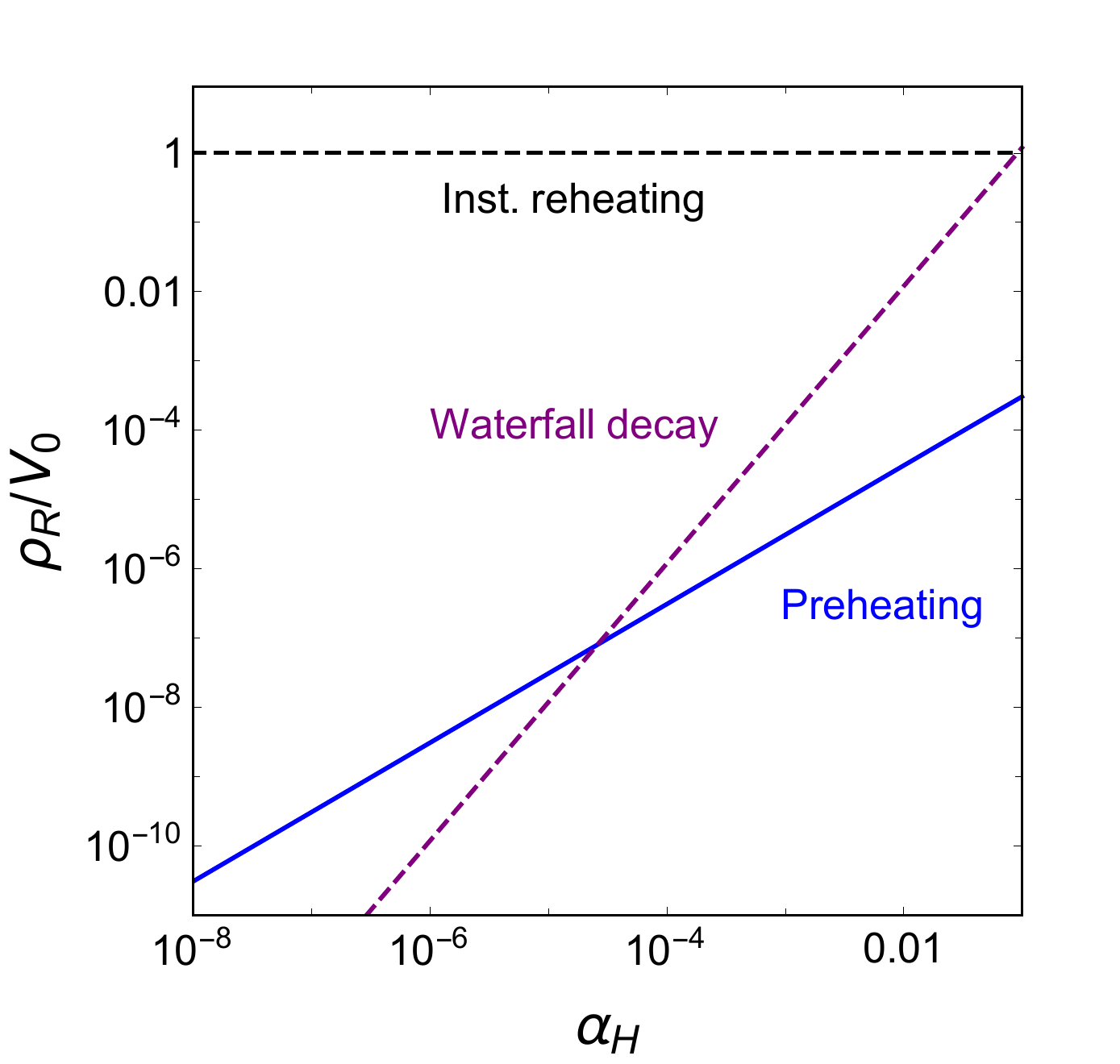}
\end{center}
\caption{$\rho_R/V_0$ as a function of $\alpha_H=\kappa_1/\lambda_\chi$, from preheating in blue solid line, and its maximum value from decay of waterfall field  in purple dashed line. We took $m_1=100\,{\rm H}_I$.  Black dashed lines correspond to the instantaneous reheating. }
\label{fig:comp}
\end{figure}

In Fig.~\ref{fig:comp}, we draw  the fraction of the radiation energy in the waterfall energy density at the end of inflation, $\rho_R/V_0$, as a function of $\alpha_H=\kappa_1/\lambda_\chi$, for preheating in the blue solid line and for the decay of the waterfall field in the purple dashed line. For comparison, we also show the case for instantaneous reheating, $\rho_R/V_0=1$, in the black dashed line.
Then, for $m_1=100 H_I$, we get $R_{\rm rh}\gtrsim 1$ for $\alpha_H\gtrsim 2.6\times 10^{-5}$, so the decay of the waterfall field is dominant for reheating.

For $R_{\rm rh}\gtrsim 1$ and $A\ll v\ll 1$, we can approximate eq.~(\ref{rad}) to
\bea
\rho_R\simeq \frac{4}{5} (\Gamma_{\chi_1}M_P)^2 v^{-1}=\frac{\pi^2}{30} g_* T^4,
\eea
and rewrite the scale factor in eq.~(\ref{scalefactor}) as a function of the radiation temperature $T$ as
\bea
\bigg(\frac{a}{a_*}\bigg)^3\simeq \bigg(\frac{v}{A}\bigg)^2=\bigg(\frac{24\Gamma^2_{\chi_1}M^2_P}{g_* \pi^2 A T^4}\bigg)^2. \label{scaleRH}
\eea
So, choosing $T=T_{\rm RH}$ in the above result, we can take into account the red-shift factor from the waterfall transition to the reheating completion.

\begin{figure}[!t]
\begin{center}
\includegraphics[width=0.325\textwidth,clip]{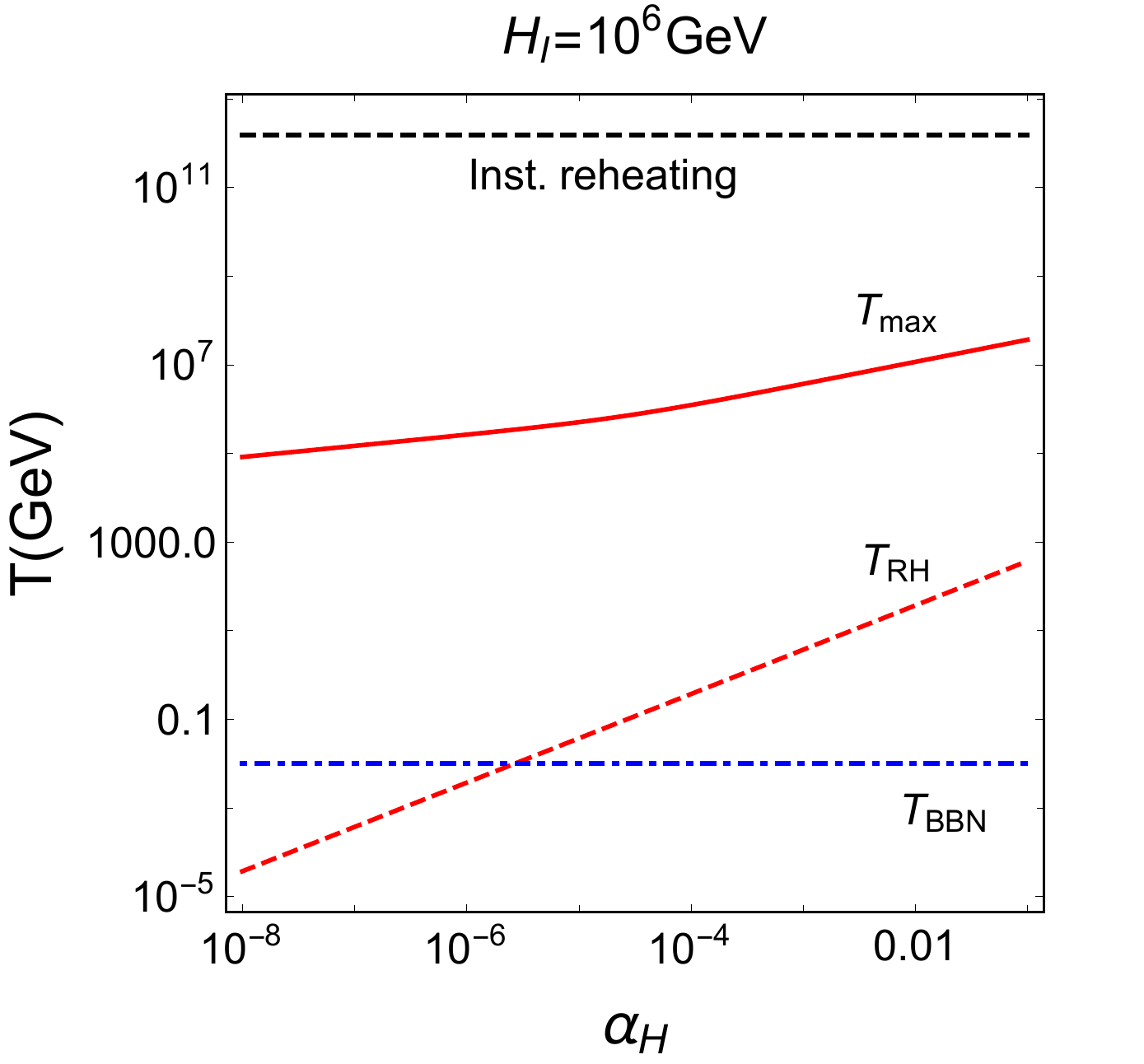}  \includegraphics[width=0.325\textwidth,clip]{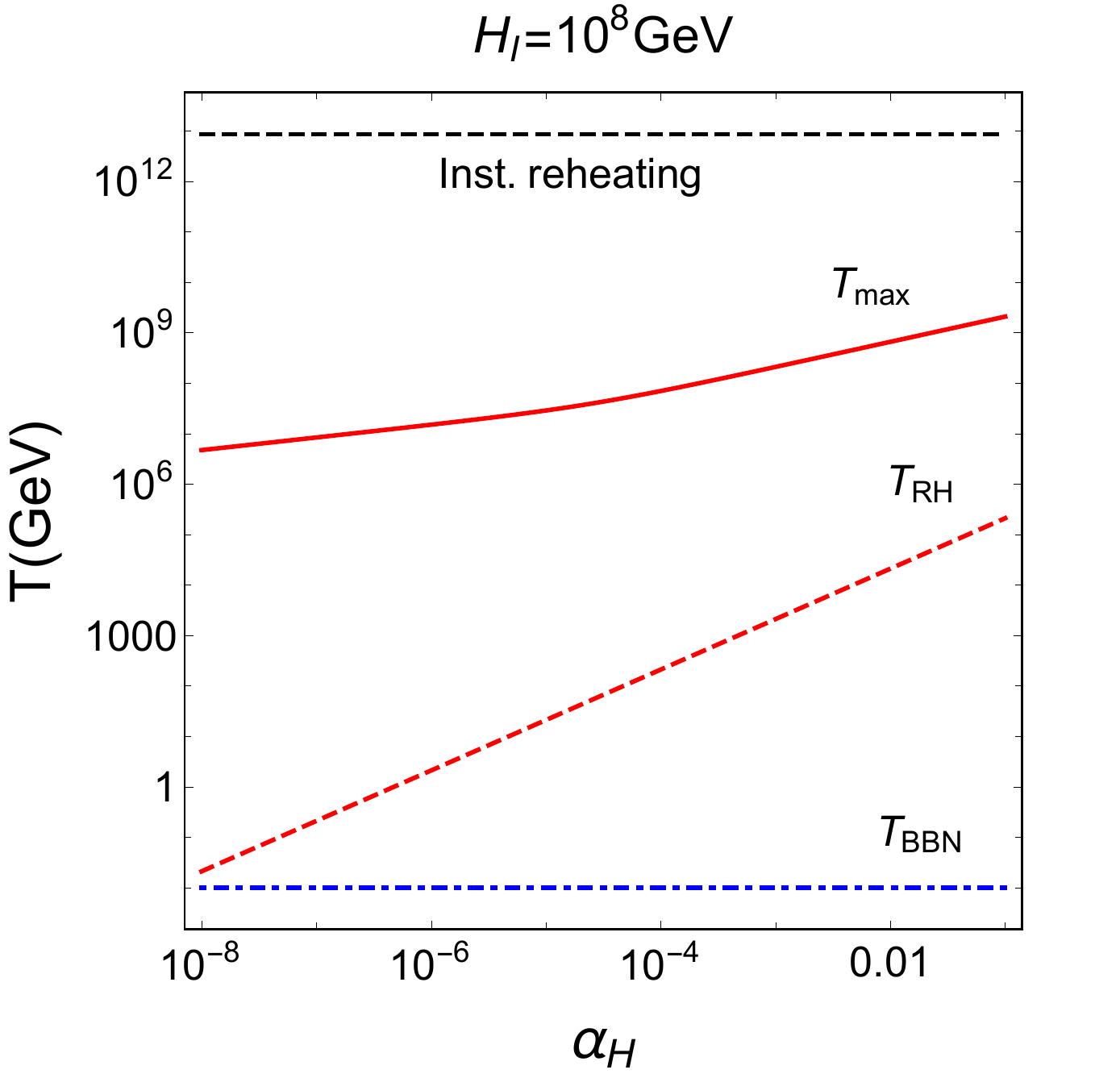} 
\includegraphics[width=0.325\textwidth,clip]{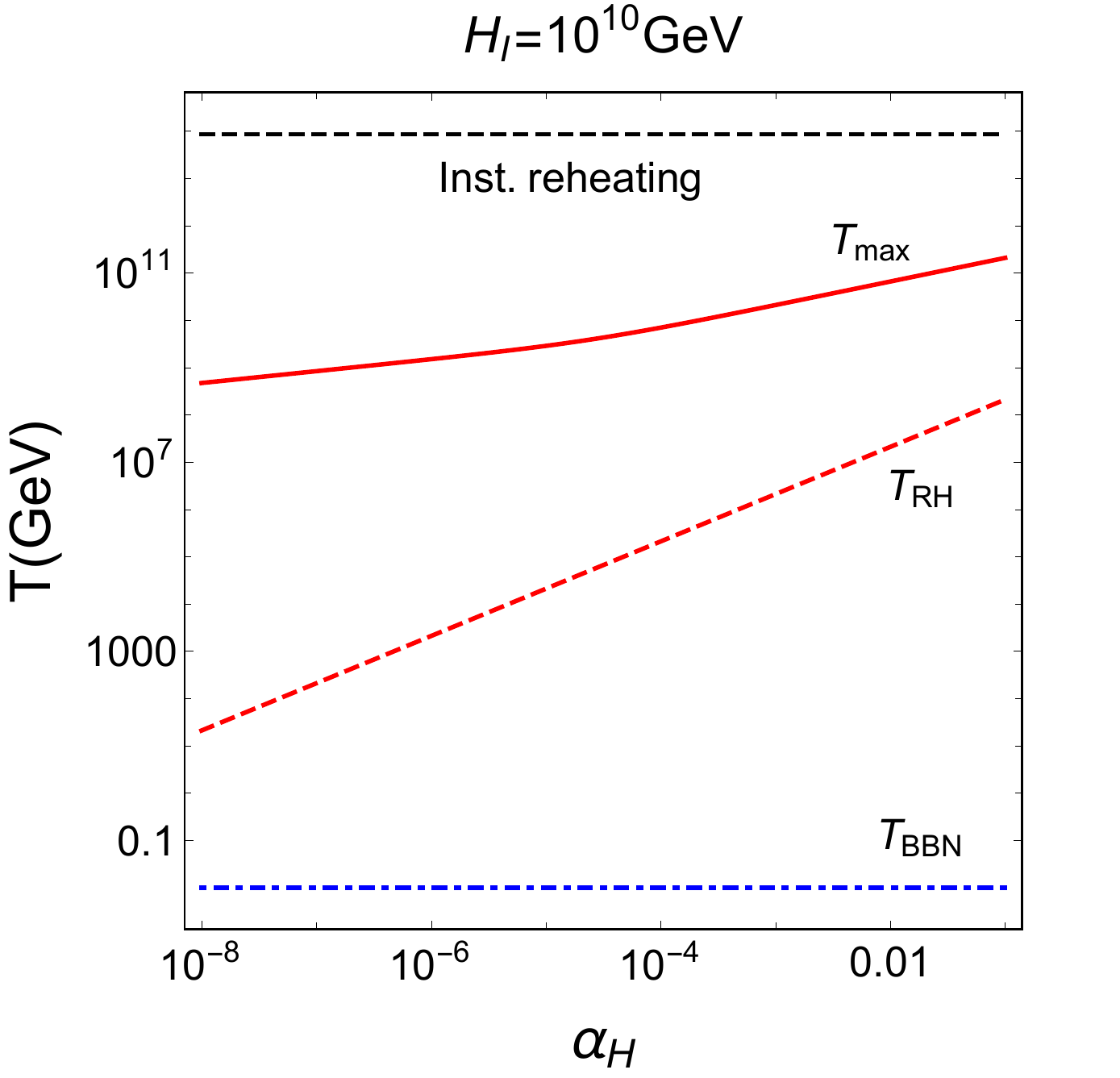} 
\end{center}
\caption{Maximum and reheating temperatures as a function of $\alpha_H=\kappa_1/\lambda_\chi$ in red solid and dashed lines, respectively, for $H_I=10^6, 10^8, 10^{10}\,{\rm GeV}$ from left to right.
We took $m_{\psi,0}=0$ and $m_1=100\,{\rm H}_I$.  Black dashed lines correspond to the instantaneous reheating and blue dot-dashed lines correspond to $T_{\rm RH}=10\,{\rm MeV}$, which is the lower bound from BBN. }
\label{fig:temp}
\end{figure}	

In Fig.~\ref{fig:temp}, we show the maximum and reheating temperatures as a function of $\alpha_H=\kappa_1/\lambda_\chi$ in red solid and dashed lines, respectively. We chose $H_I=10^6, 10^8, 10^{10}\,{\rm GeV}$ from left to right. The lower bound on the reheating temperature for BBN is set to $T_{\rm BBN}=10\,{\rm MeV}$, given in the blue dashed lines, and the black dashed lines correspond to instantaneous reheating, for comparison. Consequently, for a relatively small Hubble scale as in the left-most plot, the waterfall coupling to the Higgs field should be sizable for a successful reheating, for instance, $\alpha_H\gtrsim 5\times 10^{-6}$ for $H_I=10^6\,{\rm GeV}$. However, for a high Hubble scale, as in the middle and right-most plots, small waterfall couplings can be compatible with BBN, for which the preheating effect as shown in Fig.~\ref{fig:comp} becomes important for reheating.

\subsection{Dark matter production}

If the $Z'_2$ symmetry for the waterfall field $\chi_2$ is unbroken in the vacuum, the waterfall field $\chi_2$ can be a dark matter candidate. Dark matter can be produced from preheating, but there is no production from the decay of the waterfall field $\chi_1$ due to the kinematic blocking, namely, $m_{{\tilde \chi}_1}<2m_{\chi_2}$, from eqs.~(\ref{chi1mass}) and (\ref{chi2mass}).

From the result in eq.~(\ref{preheating}), preheating leads to the number density for dark matter $X=\chi_2$ as
\bea
n_X(t_*)=  10^{-3} m^3_1 f(\alpha_X,1.3)/\alpha_X, \label{dmnumber}
\eea
with $\alpha_X={\bar\lambda}_\chi/\lambda_\chi$.
Then, the number density produced from preheating becomes red-shifted at the reheating temperature, as follows,
\bea
n^{\rm pre}_X(T_{\rm RH})=\bigg(\frac{a_*}{a(t_{\rm RH})}\bigg)^3 n_X(t_*)=\bigg(\frac{g_* \pi^2 A T^4_{\rm RH}}{24\Gamma^2_{\chi_1}M^2_P}\bigg)^2 n_X(t_*) \label{dmpre}
\eea
where eq.~(\ref{scaleRH}) at the reheating temperature is used in the second equality. 
As a result, the relic density for dark matter $\chi_2$ at present is given by
\bea
\Omega_{\rm DM} h^2&=& \frac{m_{\chi_2}n_X(t_0)}{\rho_c/h^2}  \nonumber \\
&=&5.9\times 10^6\bigg(\frac{n^{\rm pre}_X(T_{\rm RH})}{T^3_{\rm RH}} \bigg) \bigg(\frac{m_{\chi_2}}{1\,{\rm GeV}} \bigg).
\eea

\begin{figure}[!t]
\begin{center}
\includegraphics[width=0.325\textwidth,clip]{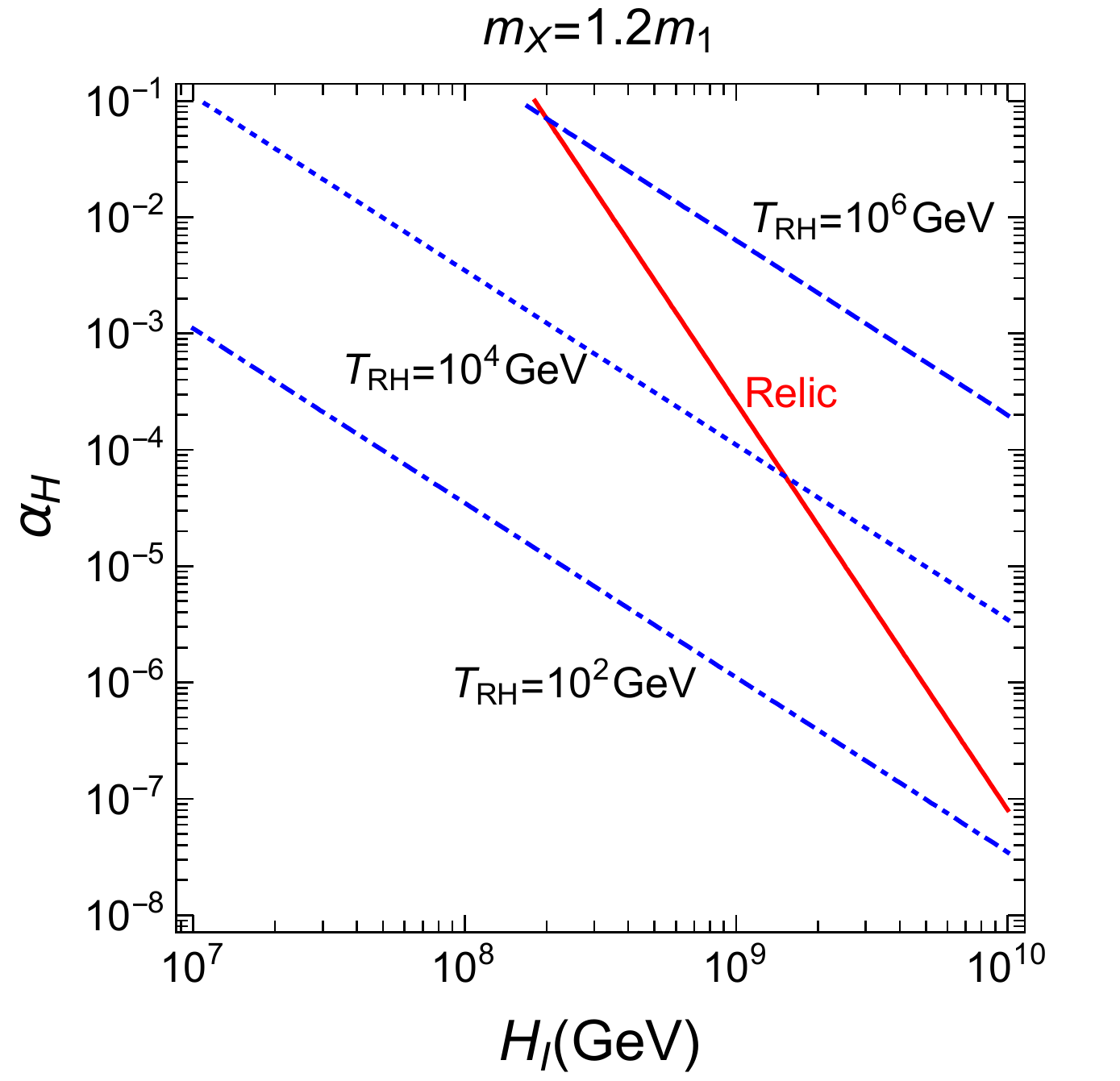}  \includegraphics[width=0.325\textwidth,clip]{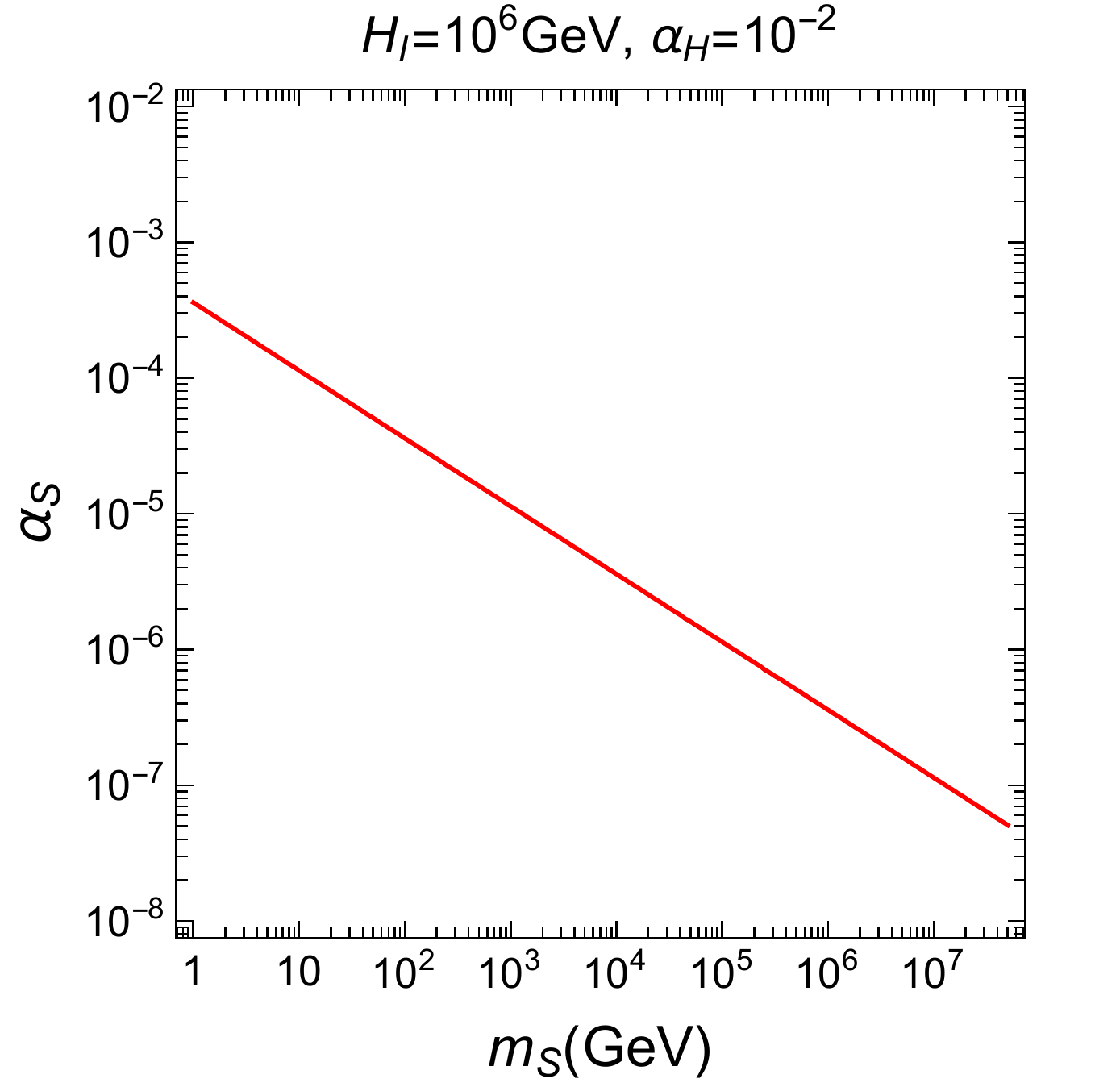}  \includegraphics[width=0.325\textwidth,clip]{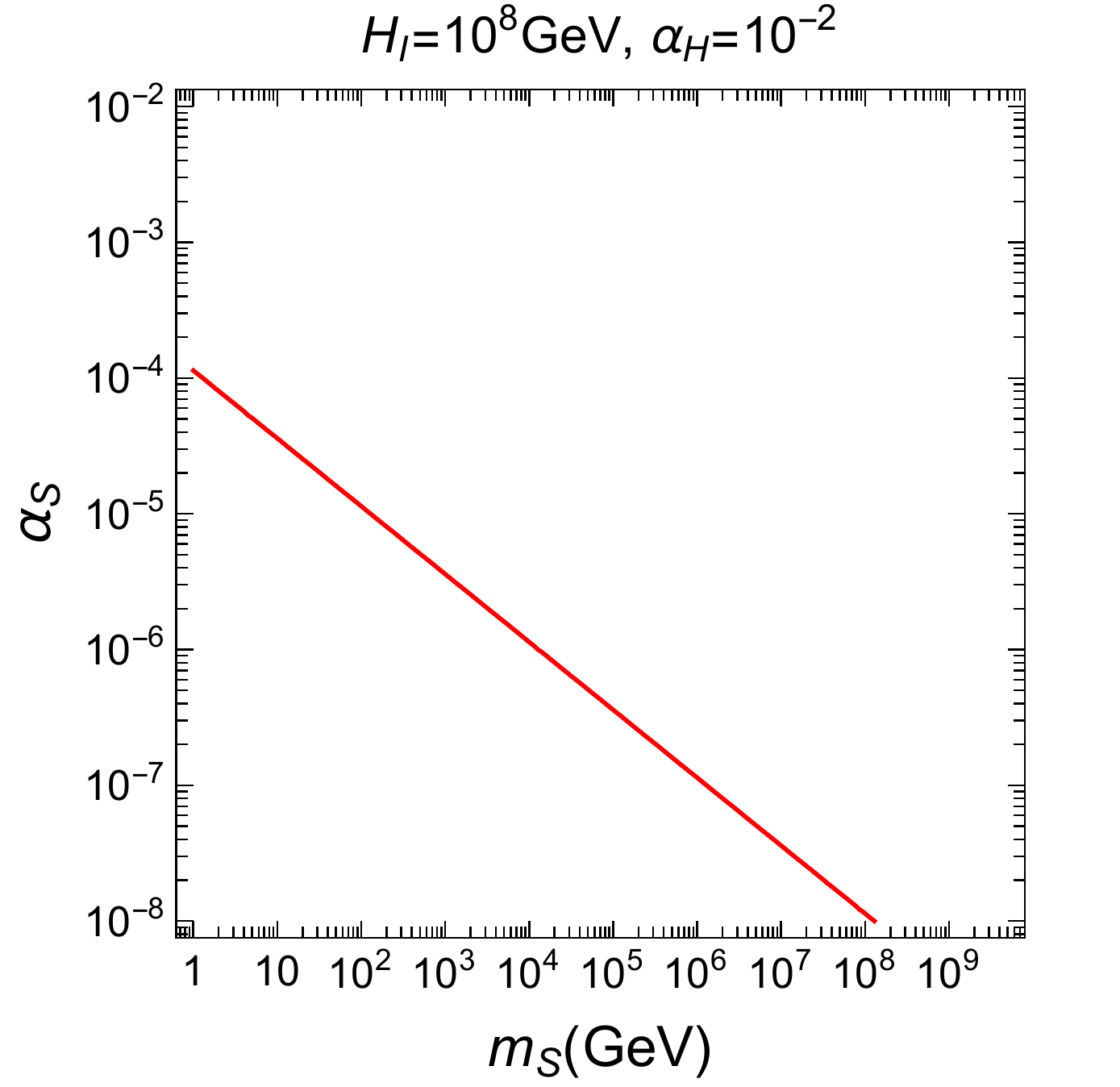} 
\end{center}
\caption{Parameter space satisfying the relic density in red lines. (Left) $H_I$ vs $\alpha_H$ with the relic density from dark matter $\chi_2$. We chose $m_X=1.2m_1$ and drew the contours with reheating temperature, $T_{\rm RH}=10^2, 10^4, 10^6\,{\rm GeV}$, in blue dashed, dotted and dot-dashed lines, respectively. The results are insensitive to $\alpha_X={\bar\lambda}_\chi/\lambda_\chi$ as far as $\alpha_X\lesssim \alpha_H$. (Middle, Right) $m_S$ vs $\alpha_S$ with the relic density from dark matter $S$. We chose $\alpha_H=10^{-2}$ for both plots, but $H_I=10^6, 10^8\,{\rm GeV}$ in the middle and right plots in order.  We took $m_1=100 H_I$ for all the plots. }
\label{fig:dm}
\end{figure}

In the left-most plot of Fig.~\ref{fig:dm}, we show the parameter space for $H_I$ vs $\alpha_H=\kappa_1/\lambda_\chi$, satisfying the relic density for dark matter $\chi_2$ in red line. We took the dark matter mass to $m_X=1.2m_1$ and $m_1=100\,H_I$. The results are insensitive to the dark matter coupling, $\alpha_X$, as far as $\alpha_X\lesssim \alpha_H$, because the numerical formula for the number density of dark matter in eq.~(\ref{dmnumber}) does not depend much on $\alpha_X$.  We show the contours of the reheating temperature, $T_{\rm RH}=10^2, 10^4, 10^6\,{\rm GeV}$, in blue dashed, dotted and dot-dashed lines, respectively. 
In this case, we find that the correct relic density can be achieved for $\alpha_H=0.1-10^{-7}$ and $H_I=2\times 10^8-10^{10}\,{\rm GeV}$. Thus, we need a relatively high reheating temperature for the relic density.

Suppose that there is a real scalar dark matter $S$, which does not participate in the waterfall transition, transforms by $S\to -S$ under the $Z'_2$ symmetry and takes the following effective squared mass,
\bea
m^2_{S,{\rm eff}}=m^2_{S,0} + \lambda_S \chi^2_1 (t)
\eea
where $m^2_{S,0}$ is the squared bare mass and $\lambda_S$ is the coupling between $S$ and the waterfall field $\chi_1$.
Then, $S$ particles can be produced not only from preheating but also from the decay of the waterfall field $\chi_1$. 
The Boltzmann equation for dark matter density after preheating is given by
\bea
{\dot n}_S + 3H n_S =2B_{\chi_1}\Gamma_{\chi_1} \frac{\rho_{\chi_1}}{m_{\chi_1}}
\eea 
where $B_{\chi_1}$ is the decay branching ratio for $\chi_1\to SS$. Then, solving the above Boltzmann equation with eq.~(\ref{RH}), we obtain the number density at the reheating temperature from the decay of the waterfall field $\chi_1$ \cite{decay} as
\bea
n^{\rm decay}_S(T_{\rm RH}) = \frac{g_*(T_{\rm RH})\pi^2 B_{\chi_1}}{18m_{\chi_1}}\, T^4_{\rm RH}.  \label{dmdecay}
\eea
As a result, we obtain the relic density for the scalar dark matter $S$ at present, as follows,
\bea
\Omega_{\rm DM} h^2&=& \frac{m_S n_S(t_0)}{\rho_c/h^2}  \nonumber \\
&=&5.9\times 10^6\bigg(\frac{n_S(T_{\rm RH})}{T^3_{\rm RH}} \bigg) \bigg(\frac{m_S}{1\,{\rm GeV}} \bigg)
\eea
where the total number density for dark matter is given by the sum of both the preheating and the decay contributions,
\bea
n_S(T_{\rm RH})= n^{\rm pre}_S(T_{\rm RH})+ n^{\rm decay}_S(T_{\rm RH}).
\eea
Here,  $ n^{\rm pre}_S(T_{\rm RH})$ is the preheating contribution, obtained from eqs.~(\ref{dmnumber}) and (\ref{dmpre}) with $\alpha_X$ being replaced by $\alpha_S$.

In the middle and right-most plots of Fig.~\ref{fig:dm}, we present the parameter space for $m_S$ vs $\alpha_S$, satisfying the correct relic density for $S$ in the case with $H_I=10^6, 10^8\,{\rm GeV}$, respectively. We chose $\alpha_H=10^{-2}$ and $m_1=100 H_I$ for both plots. As a result, we find that dark matter with masses up to $m_S=1-10^8\,{\rm GeV}$ can be produced with a correct relic density for $10^{-4}\gtrsim \alpha_S\gtrsim 10^{-8}$.

\section{Conclusions}

We have explored the inflation and the post-inflationary evolution of the universe in the hybrid inflation with a pNGB boson inflaton and two waterfall scalar fields. The couplings between the inflaton and the waterfall fields are introduced for the graceful exit, but the $Z_2$ symmetry for the twin waterfall fields ensure that the quadratic divergent loop corrections to the inflaton potential are cancelled, so the inflationary predictions of the hybrid inflation are maintained at the quantum level for a wide range of the parameter space. The $Z_2$ symmetry is broken spontaneously in the vacuum when the inflaton and the waterfall fields settle down to the local minimum of the potential, but the domain walls could be formed only beyond the horizon of our universe during inflation, leading to no observable constraints. 

We showed that the $Z_2$ invariant couplings between the waterfall fields and the SM Higgs are responsible for the reheating process via preheating and perturbative decay of the waterfall field. Thus, we achieved the post-inflationary universe with a sufficiently large reheating temperature being compatible with BBN, depending on the Higgs coupling to the waterfall field. 
On the other hand, the inflaton receives a large mass in the $Z_2$ breaking minimum, with a large VEV of the waterfall field being larger than the VEV of the other waterfall field, so it settles down to the minimum of the inflaton potential rapidly without changing the results on reheating with the waterfall fields. 

In the presence of an extra $Z'_2$ symmetry, one of the waterfall fields or another singlet scalar field becomes a dark matter candidate. Although the mass of the waterfall dark matter is constrained to be comparable to the other waterfall field by the $Z_2$ symmetry, we showed that the correct relic abundance for dark matter can be produced via preheating.

\section*{Acknowledgments}

We thank Seong Chan Park for interesting discussion on the subject. 
The work is supported in part by Basic Science Research Program through the National
Research Foundation of Korea (NRF) funded by the Ministry of Education, Science and
Technology (NRF-2022R1A2C2003567 and NRF-2021R1A4A2001897).

\def\theequation{A.\arabic{equation}}

\setcounter{equation}{0}

\vskip0.8cm
\noindent
{\Large \bf Appendix A: Preheating during the waterfall transition}

We consider preheating and particle production in the case of time-dependent masses for scalar fields by following the general formalism in Refs.~\cite{particle-prod,Mukhanov:2007zz} and review the analytic solutions for the mode number density of produced particles for a class of models with the waterfall transition discussed in the text. 

The Lagrangian for a scalar field $\phi$  in curved background is given by
\bea
{\cal L}_S = \sqrt{-g} \bigg[ \frac{1}{2} (\partial_\mu\phi)^2 -\frac{1}{2} m^2_\phi \phi^2 -\frac{1}{2} \zeta \phi^2 R \bigg].
\eea
For the FRW background with flat space, the above Lagrangian becomes
\bea
{\cal L}_S = a^3 \bigg[\frac{1}{2} (\dot\phi)^2-\frac{1}{2}a^{-2} (\partial_i\phi)^2 -\frac{1}{2} m^2_\phi \phi^2 -3\zeta\phi^2 \Big(\frac{\ddot{a}}{a} +\frac{\dot{a}^2}{a^2}\Big) \bigg].
\eea
As a result, the equation of motion for the scalar field is
\bea
\frac{d}{dt}\Big(a^3{\dot\phi}\Big)-a\, \partial^2_i\phi+a^3\, m^2_\phi \phi +6a^3\,\zeta\phi \Big(\frac{\ddot{a}}{a} +\frac{\dot{a}^2}{a^2}\Big)=0,
\eea
namely,
\bea
{\ddot\phi} +3 H {\dot \phi} -\frac{1}{a^2}\partial^2_i\phi +m^2_\phi \phi +6\zeta\phi \Big(\frac{\ddot{a}}{a} +\frac{\dot{a}^2}{a^2}\Big)=0.
\eea
Then, introducing the conformal time by $\eta=\int dt/a(t)$, the above equation becomes
\bea
\frac{1}{a} \frac{d}{d\eta} \Big(a^2 \frac{d\phi}{d\eta} \Big) -a\, \partial^2_i\phi+a^3\,m^2_\phi \phi +6\zeta\phi\, \frac{d^2a}{d\eta^2} =0,
\eea
thus
\bea
\phi^{\prime\prime} + \frac{2a'}{a}\, \phi' -\partial^2_i\phi+a^2\, m^2_\phi \phi +6\zeta\phi\,\frac{a^{\prime\prime}}{a} =0.
\eea
Therefore, making the field redefinition with $\varphi=a\,\phi$, we get 
\bea
\varphi^{\prime\prime} -\partial^2_i\varphi+\Big(m^2_\phi a^2+(6\zeta-1)\,\frac{a^{\prime\prime}}{a}  \Big)\varphi=0. \label{spert0}
\eea

We take the general solution to eq.~(\ref{spert0}) by
\bea
\varphi ({\vec x},\eta)= \int \frac{d^3 k}{(2\pi)^{3/2}}\, e^{i {\vec k}\cdot {\vec x}}\, \Big(v(k,\eta)\,a(k)+ v^*(-k,\eta)\,a^\dagger(-k)\Big) \label{ssol}
\eea
where the ladder operators satisfy 
\bea
[a(k), a^\dagger(k')]  =  \delta^3({\vec k}-{\vec k}').
\eea
Here, $v(k,\eta)$ satisfies the following equation,
\bea
v^{\prime\prime} +\Big(  \omega^2_k+(6\zeta-1)\,\frac{a^{\prime\prime}}{a}   \Big) v=0 \label{seq}
\eea
with $\omega^2_k=k^2+m^2_{\psi,{\rm eff}} a^2$, and subject to the consistency condition for the canonical quantization as
\bea
v(k,\eta) v^{*\prime}(k,\eta)-v'(k,\eta) v^*(k,\eta)=i.  \label{wronsk}
\eea

In the case with $\zeta=\frac{1}{6}$, from eq.~(\ref{ssol}), we get the Hamiltonian, 
\bea
H&=&\int d^3k \, \bigg( \frac{1}{2} (v' \,a + v^{*\prime}\, a^\dagger)^2+\frac{1}{2} (m^2_\phi a^2+ k^2) (v\,a+v^* a^\dagger)^2 \bigg) \nonumber \\
&=&\frac{1}{2} \int d^3k\, \bigg(E_k (a^\dagger a +a a^\dagger )+ F_k \, a^2 + F^*_k \,(a^\dagger)^2 \bigg). \label{Ham}
\eea
where
\bea
E_k &=&  |v'|^2 +\omega^2_k|v|^2, \\
F_k &=&(v')^2+\omega^2_k v^2.
\eea
We note from eq.~(\ref{wronsk}) that $|F_k|^2-E^2_k=-\omega^2_k$.
Then, performing the Bogolyubov transformation with
\bea
{\hat a}(k,\eta) &=& \alpha(k,\eta) a(k) -\beta(k,\eta) a^\dagger(k), \\
{\hat a}^\dagger(k,\eta) &=& -\beta^*(k,\eta) a(k) +\alpha^*(k,\eta) a^\dagger(k)
\eea
with $|\alpha|^2-|\beta|^2=1$,
we can diagonalize the Hamiltonian in eq.~(\ref{Ham}) to
\bea
H=\frac{1}{2} \int d^3k\, \omega_k\, \Big[{\hat a}^\dagger(k,\eta) {\hat a}(k,\eta) +{\hat a}(k,\eta) {\hat a}^\dagger(k,\eta) \Big],
\eea
provided that
\bea
\frac{\alpha}{\beta}=\frac{F^*_k}{\omega_k-E_k}, \quad |\beta|^2= \frac{(\omega_k-E_k)^2}{|F_k|^2-(\omega_k-E_k)^2}=-\frac{1}{2}+\frac{E_k}{2\omega_k}.
\eea

We define a quasi-particle vacuum $|0_\eta\rangle$ such that ${\hat a}|0_\eta\rangle=0$.
Then, the total number density of produced particles up to time $\eta$ is given by
\bea
n(\eta) = \langle 0_\eta| N |0_\eta\rangle= \frac{1}{2\pi^2a^3(\eta)}\, \int^\infty_0 dk\, k^2 |\beta|^2
\eea
where the number operator for particles, $N=a^\dagger a$ with $a= \alpha^* {\hat a} +\beta\, {\hat a}^\dagger$, is used and
\bea
|\beta|^2 &=&-\frac{1}{2} +\frac{E_k}{2\omega_k} \nonumber \\
&=&- \frac{1}{2} +\frac{1}{2\omega_k} \Big(|v'|^2 +\omega^2_k |v|^2 \Big).  \label{betas}
\eea

Taking the initial conditions for the Minkowski vacuum at $\eta=0$,
\bea
v'(0) &=& -i\omega_k\, v(0), \\
v(0) &=& \frac{1}{\sqrt{2\omega_k}},
\eea
which is consistent with eq.~(\ref{wronsk}) and the minimum zero-point energy,
the quantity $|\beta|^2$ at $\eta=0$ vanishes,
\bea
|\beta|^2= -\frac{1}{2} + \frac{1}{2\omega_k} \Big(|v'(0)|^2 +\omega^2_k |v(0)|^2 \Big)=0.
\eea

We consider the following form of the time-dependent effective mass for the scalar field \cite{tanh},
\bea
(m^2\,a^2)(\eta)=  m^2 (\alpha+\rho\tanh\lambda\eta+\gamma\tanh^2\lambda  \eta). \label{mass0}
\eea
In this case, the effective mass squared changes from $m^2(\alpha-\rho+\gamma)$ to $m^2(\alpha+\rho+\gamma)$.

Now making a change of variables with $\xi=\frac{1}{2}(1+\tanh\lambda \eta)$ and  the field redefinition with
\bea
v(\eta)= \xi^{c_1} (1-\xi)^{c_2} u(\xi),
\eea
with
\bea
c_1 &=& \mp \frac{i\omega_1}{2\lambda}, \\
c_2 &=& \pm  \frac{i\omega_2}{2\lambda},
\eea
where $\omega_1=\sqrt{k^2+m^2(\alpha+\gamma-\rho)}$ and $\omega_2=\sqrt{k^2+m^2(\alpha+\gamma+\rho)}$, 
we can rewrite eq.~(\ref{seq}) with $\zeta=\frac{1}{6}$ into the differential equation for hypergeometric function,
\bea
\xi(1-\xi) \frac{d^2 u}{d\xi^2} + \Big[c-(a+b+1)\xi \Big] \frac{d u}{d\xi} -a\,b \, u =0
\eea
with 
\bea
c&=& 1- \frac{i\omega_1}{\lambda}, \\
a&=& \frac{1}{2} +i \Big( \frac{\omega_-}{\lambda}+\delta\Big), \\
b&=& \frac{1}{2} +i \Big( \frac{\omega_-}{\lambda}-\delta\Big)
\eea
where $\omega_\pm=(\omega_2\pm\omega_1)/2$ and
\bea
\delta\equiv \sqrt{\frac{\gamma m^2}{\lambda^2}-\frac{1}{4}}.
\eea

Under the boundary condition with $v(\eta)=\frac{1}{\sqrt{2\omega_1}}\, e^{-i\omega_1\eta}$ at $\eta=-\infty$, we choose $c_1=-\frac{i\omega_1}{2\lambda}$  and $c_2=\frac{i\omega_2}{2\lambda}$, so the solution takes
\bea
v(\eta)=A\,  \xi^{c_1} (1-\xi)^{c_2} \, {}_2F_1(a,b,c;\xi)
\eea
with
\bea
A&=&\frac{1}{\sqrt{2\omega_1}}.
\eea
Here, we note that ${}_2F_1(a,b,c;\xi)\rightarrow 1$ for $\xi\rightarrow 0$ or $\eta\rightarrow-\infty$, and $\xi\approx e^{2\lambda\eta}$ for $\eta\rightarrow -\infty$  and $1-\xi\approx e^{-2\lambda\eta}$ $\eta\rightarrow +\infty$. 

In order to find the number of produced scalar particles at $\eta\rightarrow +\infty$, using the following identity for the hypergeometric function,
\bea
{}_2F_1(a,b,c;\xi)&=&B_1\, \cdot {}_2F_1(a,b,a+b-c+1;1-\xi) \nonumber \\
&&+ B_2\cdot (1-\xi)^{c-a-b} \,{}_2F_1(c-a,c-b,c-a-b+1;1-\xi)
\eea
with $c-a-b=-2c_2$ and
\bea
B_1&=& \frac{\Gamma(c)\Gamma(c-a-b)}{\Gamma(c-a)\Gamma(c-b)}, \\
B_2&=&\frac{\Gamma(c)\Gamma(a+b-c)}{\Gamma(a)\Gamma(b)},
\eea
we find the asymptotic form for $v$ for $\xi\rightarrow 1$ as
\bea
v\approx A\, \Big(B_1 \,e^{-i\omega_2 \eta} + B_2\, e^{i\omega_2\eta} \Big),
\eea
so
\bea
v'\approx -i\omega_2 A \, \Big(B_1 \,e^{-i\omega_2 \eta} - B_2\, e^{i\omega_2\eta} \Big).
\eea
Therefore, from eq.~(\ref{betas}), for $\xi\rightarrow 1$, we get
\bea
|\beta|^2&\approx& -\frac{1}{2} +\frac{\omega_2}{2\omega_1} \, (|B_1|^2+|B_2|^2) \nonumber \\
&=& -\frac{1}{2} +\frac{\omega_2}{2\omega_1} \left|\frac{\Gamma\Big(1-\frac{i\omega_1}{\lambda}\Big)\Gamma\Big(-\frac{i\omega_2}{\lambda}\Big)}{\Gamma\Big(\frac{1}{2}-i\Big(\frac{\omega_+}{\lambda}+\delta\Big)\Big)\Gamma\Big(\frac{1}{2}-i\Big(\frac{\omega_+}{\lambda}-\delta\Big)\Big)} \right|^2 \nonumber \\
&&+ \frac{\omega_2}{2\omega_1}\left|\frac{\Gamma\Big(1-\frac{i\omega_1}{\lambda}\Big)\Gamma\Big(\frac{i\omega_2}{\lambda}\Big)}{\Gamma\Big(\frac{1}{2}+i\Big(\frac{\omega_-}{\lambda}+\delta\Big)\Big)\Gamma\Big(\frac{1}{2}+i\Big(\frac{\omega_-}{\lambda}-\delta\Big)\Big)} \right|^2. \label{beta2s}
\eea

Finally, using  $\Gamma(1+x)=x\Gamma(x)$ and
\bea
|\Gamma(iy)|^2&=& \frac{\pi}{y \sinh(\pi y)},  \\
\Big|\Gamma\Big(\frac{1}{2}+iy\Big)\Big|^2 &=& \frac{\pi}{\cosh(\pi y)},
\eea
we obtain eq.~(\ref{beta2s}) as
\bea
|\beta|^2&\approx&- \frac{1}{2} +\frac{1}{2} \, \frac{1}{ \sinh(\pi \omega_1/\lambda)\sinh(\pi \omega_2/\lambda)}  \times \nonumber \\
&&\times\bigg[  \cosh\Big(\pi \Big(\frac{\omega_+}{\lambda}+\delta \Big)\Big)  \cosh\Big(\pi \Big(\frac{\omega_+}{\lambda}-\delta \Big)\Big) \nonumber \\
&&\quad +\cosh\Big(\pi \Big(\frac{\omega_-}{\lambda}+\delta \Big)\Big)  \cosh\Big(\pi \Big(\frac{\omega_-}{\lambda}-\delta \Big)   \Big) \bigg] \nonumber \\
&=&\frac{ \cosh(2\pi \omega_-/\lambda)+\cosh(2\pi\delta) }{ 2\sinh(\pi \omega_1/\lambda)\sinh(\pi \omega_2/\lambda)}. \label{betafinal}
\eea

For instance, for $\alpha=\gamma=\frac{1}{2}\rho=\frac{1}{4}$, the effective scalar mass in eq.~(\ref{mass0}) becomes
\bea
(m^2 a^2)(\eta) = \frac{1}{4} m^2 (1+\tanh\lambda\eta)^2.
\eea
In this case, the above result in eq.~(\ref{betafinal}) becomes
\bea
|\beta|^2&\approx&\frac{ \cosh(2\pi \omega_-/\lambda)+\cosh\Big(\pi\sqrt{\frac{m^2}{\lambda^2}-1}\Big) }{ 2\sinh(\pi \omega_1/\lambda)\sinh(\pi \omega_2/\lambda)}. \label{beta}
\eea

On the other hand, for $\alpha=\rho=\frac{1}{2}$ and $\gamma=0$, for which the effective scalar mass in eq.~(\ref{mass0}) becomes
\bea
(m^2 a^2)(\eta) = \frac{1}{2} m^2 (1+\tanh\lambda\eta),
\eea
the above result in eq.~(\ref{betafinal}) becomes
\bea
|\beta|^2&\approx&\frac{ \sinh^2(\pi \omega_-/\lambda) }{ \sinh(\pi \omega_1/\lambda)\sinh(\pi \omega_2/\lambda)}.
\eea

\end{document}